\documentclass[journal=jacsat,manuscript=article]{achemso}

\usepackage[version=3]{mhchem} 
\usepackage{booktabs}

\author{Qi Ou}
\affiliation{AI for Science Institute, Beijing 100080, P.R.~China}
\email{ou_qi@163.com}
\author{Ping Tuo}
\affiliation{AI for Science Institute, Beijing 100080, P.R.~China}
\author{Wenfei Li}
\affiliation{AI for Science Institute, Beijing 100080, P.R.~China}
\author{Xiaoxu Wang}
\affiliation{DP Technology, Beijing 100080, P.R.~China}
\author{Yixiao Chen}
\affiliation{Program in Applied and Computational Mathematics, Princeton University, Princeton, NJ 08544, USA}
\author{Linfeng Zhang}
\affiliation{AI for Science Institute, Beijing 100080, P.R.~China}
\alsoaffiliation{DP Technology, Beijing 100080, P.R.~China}
\email{linfeng.zhang.zlf@gmail.com}

\title[An \textsf{achemso} demo]
  {DeePKS Model for Halide Perovskites with the Accuracy of Hybrid Functional}

\begin{document}








\begin{abstract}
 Accurate prediction for the electronic structure properties of halide perovskites plays a significant role in the design of highly efficient and stable solar cells. While density functional theory (DFT) within the generalized gradient approximation (GGA) offers reliable prediction in terms of lattice constants and potential energy surface for halide perovskites, it severely underestimates the band gap due to the lack of non-local exact exchange term, which exists in computationally expensive hybrid functionals. In this work, a universal Deep Kohn-Sham (DeePKS) model based on neural network is trained so as to enable electronic structure calculations with the accuracy of hybrid functional HSE06 and the efficiency comparable to GGA functional, for a plethora of halide perovskites, i.e., ABX$_3$ (A=FA, MA, Cs; B=Sn, Pb; X=Cl, Br, I). Forces, band gaps, and density of states (DOS) predicted by our DeePKS model for all aforementioned perovskites are in good agreement with the HSE06 results, with significantly improved efficiency. In addition, even though the spin-orbit coupling (SOC) effect has not been taken into consideration during the training process, DeePKS+SOC offers highly consistent band gap and DOS as compared to HSE06+SOC for Pb-containing systems. We believe such DeePKS model can be readily applied for an accurate yet efficient prediction of various properties for the family of halide perovskites.   
\end{abstract}

\section{Introduction}
Owing to the unsustainable nature of the fossil-based-energy sources, new forms of 
energy sources that are both renewable and environmentally friendly have been 
comprehensively investigated world-wide.\cite{energy1, energy2, energy3, energy4, energy5} Solar cell, or photovoltaic, is regarded 
as one of the promising alternatives to traditional fossil fuel resources.\cite{pv1,pv2,pv3,pv4} While the silicon-based solar cells, i.e., the first- and second-generation solar cells, exhibit good performance in terms of stability and efficiency,\cite{silicon1,silicon2,silicon3,silicon4} the cost of ultra-high-pure metallic silicon somewhat hinders the application of such solar cells. Third-generation photovoltaic including organic solar cells and dye-sensitized solar cells has been developed later on, with significantly reduced processing cost.\cite{opv1,dssc1,opv2,opv3} Nevertheless, the power conversion efficiency (PCE) of the third-generation photovoltaic remains less satisfactory compared with the silicon-based modules,\cite{opv_pce1,opv_pce2,perov_pv1} which limits their chances for commercialization. 

First synthesized in 1890s,\cite{first_perov} halide perovskites (ABX$_3$, X=halogen anion) have shown immense potential to become a low-cost alternative to the currently
commercialized photovoltaic technologies owing to their desired semiconducting properties and the economic fabrication.\cite{perov_pv1,perov_pv2,perov_pv3,perov_pv4,perov_pv5} In addition to the relatively high PCE reached by halide perovskites, one of their favored traits that is rarely possessed by other conventional semiconductors is the tunable absorption edge wavelength (band gap),\cite{tunable_band1,tunable_band2,tunable_band3} which can be straightforwardly realized by varying the ratio of different halide ions.
Accurate theoretical description of the electronic structure for halide perovskites is a long-standing goal for the realistic design of photovoltaic with higher efficiency and better stability. Density functional theory (DFT) with the generalized gradient approximation (GGA) combined with D3 dispersion correction has been extensively applied in previous studies, offering a balanced trade-off between accuracy and efficiency in predicting properties such as lattice constants and potential energy surfaces for various perovskites.\cite{PhysRevB.105.064104,C9CP03240A,pbe1,pbe2,pbe3,https://doi.org/10.1002/aenm.201701136} Nevertheless, such methodology suffers a severe underestimation of the band gaps, which can only be correctly captured with either the hybrid functional that incorporates a portion of exact exchange from Hartree-Fock theory or the many-body perturbation theory such as the GW method.\cite{PhysRevB.89.155204,doi:10.1021/jz401532q,https://doi.org/10.1002/aenm.201701136,doi:10.1021/acs.jpclett.7b02648,doi:10.1021/acs.jpclett.1c03428} Unfortunately, both of these two methods are computationally much more expensive compared to GGA and even become prohibitive for large systems. 

With the development of massive computational resources and advanced machine learning (ML) algorithms, efficient yet accurate descriptions of the electronic structure for large systems have been made possible. Proposed in 2020, the Deep Kohn-Sham (DeePKS) approach introduces a general framework for generating highly accurate self-consistent energy functionals with remarkably reduced computational cost,\cite{chen2020ground,chen2020deepks,deepks_abacus} in which the energy difference between the expensive high-level method and the cheap low-level method is fitted by the neural network. While DeePKS has been comprehensively validated on molecular systems and specific condensed phase systems,\cite{chen2020deepks,deepks_abacus} its generalizability to multiple or a family of systems, though theoretically feasible, remains unexplored. Indeed, it would be essential to construct a general DeePKS model that can offer highly accurate electronic structure results with low computational cost for systems like halide perovskites, given their undeniable significance in photovoltaic and other fields.  

In this work, our goal is to establish a general DeePKS model that can compute the electronic structure properties with the accuracy of the hybrid functional HSE06 and the efficiency comparable to GGA level for various halide perovskites, i.e., ABX$_3$ (A=FA, MA, Cs; B=Sn, Pb; X=Cl, Br, I) as well as the organic-inorganic hybrid alternatives. Based on an iterative training process with 460 configurations over seven types of halide perovskites, we demonstrate that the resulting DeePKS model is capable to reproduce closely matched forces, stress, band gaps, and density of states (DOS) near the Fermi energy as compared to HSE06 for any constitution of the aforementioned ABX$_3$ as well as the hybrid ones. In addition, even though the spin-orbit coupling (SOC) effect has not been taken into consideration during the training process, our computational results show that DeePKS+SOC offers highly consistent band gap and DOS as compared to HSE06+SOC for Pb-containing systems, in which the SOC effect is necessary for a quantitatively correct description of the band gap.\cite{doi:10.1021/acs.jpcc.1c09594} The equation of state (EoS) and corresponding band gap predictions with respect to varied sizes of the unit cell are also validated. We believe such DeePKS model can be readily applied for an accurate yet efficient prediction of various properties for the family of halide perovskites and potentially benefit the discovery and design of better photovoltaic based on such materials.

\section{Methodologies and computational details}
The DeePKS algorithm for periodic systems has been previously implemented in 
an open-source DFT package ABACUS in the basis of 
numerical atomic orbitals (NAO),\cite{li_abacus,chen_abacus} of which the theoretical details can be found in 
Ref.~\citenum{deepks_abacus}. Extended labels for DeePKS training process implemented for this work are stress and band gap, which enable the converged DeePKS model to reproduce the stress and DOS near the Fermi energy given by the target method. The stress given by the DeePKS model can be calculated via:
\begin{align}
    \sigma^{\scriptsize{\textrm{DeePKS}}}_{\alpha\beta}[\{\phi_i|\omega\}] = & \sigma^{\scriptsize{\textrm{baseline}}}_{\alpha\beta}[\{\phi_i|\omega\}] - \frac{\partial E^\delta [\{\phi_i|\omega\},\omega]}{\partial \varepsilon_{\alpha\beta}} \nonumber\\
        = & \sigma^{\scriptsize{\textrm{baseline}}}_{\alpha\beta}[\{\phi_i|\omega\}] 
        - \sum_{Inlmm'} \frac{\partial E_\delta}{\partial D_{nlmm'}^I} \sum_i \frac{d}{d\varepsilon_{\alpha\beta}}[f_i\langle \phi_i | \alpha_{nlm}^I \rangle \langle \alpha_{nlm'}^I |\phi_i \rangle].
\end{align}
where $f_i$ is the occupation number of orbital $\phi_i$. Note that the stress label for DeePKS training process is of the energy unit, i.e., without dividing by the volume of the unit cell, and only the upper triangle part of the stress tensor is served as labels. For band gap labels, the band gaps of each $k$ point, i.e., the energy differences between the highest valance band (HVB) and the lowest conducting band (LCB) at each $k$ point, 
\begin{eqnarray}
\epsilon_{\textrm{g}, k} := \epsilon_{\scriptsize\textrm{LCB}, k} - \epsilon_{\scriptsize\textrm{HVB}, k},
\end{eqnarray}
are served as labels to the training process, aiming to provide an accurate description of the DOS near the Fermi energy. Total energies, forces, stresses, and band gaps of each $k$ points of the target method, are included in the loss function $L(\omega)$, and the optimization problem now becomes:
\begin{eqnarray}\label{eqn::target}
\min_\omega L(\omega), ~    
L(\omega) &=&|E^{\scriptsize{\textrm{target}}}-E^{\scriptsize{\textrm{DeePKS}}}[\{\phi_i|\omega\}]|^2 \nonumber\\
&+& 
\lambda_1 |\mathbf{F}^{\scriptsize{\textrm{target}}}-\mathbf{F}^{\scriptsize{\textrm{DeePKS}}}[\{\phi_i|\omega\}]|^2  \nonumber\\
&+&\lambda_2 |\boldsymbol{\epsilon_{\textrm{g}}}^{\scriptsize{\textrm{target}}}-\boldsymbol{\epsilon_{\textrm{g}}}^{\scriptsize{\textrm{DeePKS}}}[\{\phi_i|\omega\}]|^2 \nonumber\\
&+& \lambda_3 |\boldsymbol{\sigma}^{\scriptsize{\textrm{target}}}-\boldsymbol{\sigma}^{\scriptsize{\textrm{DeePKS}}}[\{\phi_i|\omega\}]|^2,
\end{eqnarray}
where $\lambda_1$, $\lambda_2$, and $\lambda_3$ correspond to the weighting factor of force, band gap, and stress in the loss function, are are set to be 1.0, 0.1, and 1.0 in this work, respectively. 

It should be noted that the goal of this work is to contrive a general DeePKS model for various halide perovskites instead of a single system. This poses challenges to the energy fitting process by the neural network owing to the fact that the absolute energy difference between the base and the target method contributed from each element of multiple systems might be significantly variant. To circumvent this obstacle, we apply the method of linear least squares to calculate the element-specific contribution to the absolute energy difference between the base and the target method. Specifically, the sum of squares to be minimized in our case is
\begin{eqnarray}
S = \sum_{s}[\Delta E_s - \sum_I \epsilon_I N_{I,s}]^2
\end{eqnarray}
where $\Delta E_s$ is the energy difference between the base and target method of system $s$, and $N_{I,s}$ is the number of atoms of element $I$ contained in system $s$. The minimization of $S$ gives $\epsilon_I$, i.e., the contribution of element $I$ to the energy difference between the base and target method, and the absolute energy difference resulted from $\epsilon_I$ for each system $s$, i.e., $\sum_I \epsilon_I N_{I,s}$, is then deducted from the loss function.

The training sets of the DeePKS model consist of FAPbI$_3$, 
MAPbI$_3$, CsPbI$_3$, MAPbBr$_3$, MASnCl$_3$, CsPbCl$_3$, and CsSnBr$_3$, and 
the number of configurations of each system is given in Table S2. 
Configurations of the aforementioned XPbI$_3$ family are randomly (subject to uniform 
distribution) picked from previously reported DeePMD training data,\cite{https://doi.org/10.1002/adfm.202301663} while 
those of other systems are randomly picked from corresponding DP-GEN jobs. The 
test sets consist of four systems that are different from the training sets, 
i.e., MAPbCl$_3$, MASnBr$_3$, CsSnI$_3$, and FA$_{0.125}$Cs$_{0.875}$PbI$_3$. Configurations of FA$_{0.125}$Cs$_{0.875}$PbI$_3$ are 
randomly picked from previously reported DeePMD training data,\cite{https://doi.org/10.1002/adfm.202301663} while those of 
the other three are randomly picked from corresponding DP-GEN jobs. 

The target functional applied in this work for label generation is the hybrid GGA functional HSE06. Self-consistent field (SCF) jobs for all configurations in the training test sets are carried out with the HSE06 functional and projector-augmented wave (PAW) method in the Vienna Ab initio Simulation Package (VASP).\cite{vasp1, vasp2, vasp3} The energy cutoff is set to be 500 eV, and the allowed spacing between $k$ points (``kspacing'') is set to be 0.1 Bohr$^{-1}$. Such energy cutoff has been previously validated in Refs.\citenum{https://doi.org/10.1002/adfm.202301663} and \citenum{cspbi3} for XPbI$_3$ (X=Cs, FA, MA) systems, and the kspacing value for the HSE06 jobs is slightly larger compared to that applied in Ref.\citenum{https://doi.org/10.1002/adfm.202301663} for PBE calculation (i.e., 0.16 $\mathring{\textrm{A}}^{-1}$) due to the high computational cost of the HSE06 functional. We will demonstrate in next section that the DeePKS model trained with a relatively sparse $k$-point label can successfully recover the energy, force, and DOS of the target method with a much denser $k$-point grid. During the DeePKS iterative training process, the single-point SCF calculations are performed in ABACUS\cite{chen_abacus,li_abacus} with double-zeta polarized (DZP) NAO basis,\cite{lcao1,lcao2} 100 Ry energy cutoff, 0.1 Bohr$^{-1}$ kspacing value, and the SG15 Optimized Norm-Conserving Vanderbilt (ONCV)\cite{oncv} pseudopotentials. 
The aforementioned $\epsilon_I$ of all elements involved in these systems is listed in Table S1. More details of the DeePKS training process can be found in the Supporting Information. 

\section{Results and discussion}
\subsection{Accuracy crossing-checking of the DeePKS model}
Mean absolute errors (MAE) of energies, forces, band gaps for multi-$k$ points, and stresses of the training set given by the converged DeePKS model are listed in Table S2, while those for the test set are listed in Table S3. Note that the absolute energies shift significantly between the HSE06 calculation and the DeePKS/PBE calculation due to the fact that PAW is employed for HSE06 jobs while ONCV pseudopotentials are applied in DeePKS/PBE. We exclude the influence of such ``constant'' shift by defining the mean absolute relative error (MARE) of the energy as
\begin{eqnarray}
E_{\textrm{MARE}} = \frac{1}{N}\sum_{i=1}^N |E^{\textrm{HSE06}}_i - [E^{\textrm{DeePKS/PBE}}_i - \bar{E}^{\textrm{DeePKS/PBE}} +  \bar{E}^{\textrm{HSE06}} ]|
\end{eqnarray}
It can be seen that compared to PBE results, the MAE/MARE of four labels given by DeePKS is significantly reduced, especially that of band gaps and stresses, and similar accuracy is achieved via DeePKS 
for the test sets. As shown in Table S3, for four tested systems, pronounced decrease of the MAE is observed for the band gap and stress (0.053 
eV and 0.123 eV, respectively) compared to the PBE results (0.825 eV and 
0.657 eV, respectively), while the MAE/MARE for energy and force (2.2 meV/atom, 
0.045 eV/$\mathring{\textrm{A}}$, respectively) is also decreased by around 
a factor of two. 

It should be noted that for halide perovskite systems, PBE has been validated as a reliable functional to describe the potential energy surfaces.\cite{https://doi.org/10.1002/adfm.202301663,PhysRevB.105.064104,C9CP03240A,pbe1,pbe2,pbe3,https://doi.org/10.1002/aenm.201701136} The key problem encountered by PBE is indeed the prediction of the band gap for these systems, which is essential for photovoltaic properties. HSE06, albeit computationally demanding, is able to offer significantly improved description of the band gap for perovskites. The successful recovery of the HSE06 band gap of DeePKS evinces its capability to offer a notably more accurate description of the band gap as compared to PBE. 
The reason that the stress MAE of DeePKS remains relatively large ($> 0.1$ eV) is 
attributed to the severe deviation between the PBE predicted stress and the HSE06 predicted one (with an overall MAE larger than 1.3 eV for the training set). That being said, the MAE of the stress predicted by the DeePKS model decreases remarkably as compared to that given by PBE, indicating the strength of the training process.

\subsection{Property prediction on various halide perovskite systems}
We extensively test our DeePKS model on 18 different types of organic or 
inorganic perovskites (mostly cubic phase unless otherwise specified), and the hybrid organic-inorganic 
perovskite FA$_a$Cs$_{1.0-a}$PbI$_3$ with varied $a$ values. Cell 
relaxation is performed for all non-hybrid perovskites, while ionic relaxation is performed for hybrid perovskites. All relaxation calculations are carried out with PBE functional and D3\_0
van der Waals (vdW) dispersion correction in ABACUS. DZP NAO basis set with ONCV 
pseudopotentials is applied, and the energy cutoff is set to be 100 Ry. For all cubic phase systems, the Brillouin zone is sampled with a gamma-centered $9\times9\times9$ $k$-point grid, while for all other systems, a $k$-spacing value is set to be 0.07 Bohr$^{-1}$. 
DeePKS/PBE and HSE06 single-point SCF calculations are performed on these optimized structures, with the same 
basis sets, energy cutoff as those applied in the iterative training and label generation processes, while the $k$-point settings are consistent with the aforementioned relaxtion calculations. Detailed input parameters of HSE06 (carried out in VASP) and DeePKS/PBE (carried out in ABACUS) are provided in the Supporting Information.

Total CPU times of the SCF calculations performed via HSE06, DeePKS, and PBE for these tested perovskites are given in Table S4. It can been seen that the CPU time of DeePKS is of the same order of magnitude (though twice to three times as long) as compared to that of PBE for most tested systems, while it is around two orders of magnitude faster as compared to that of HSE06 based on our input parameter and parallelization settings. In addition to the omission of constructing the Hartree-Fock exact exchange part, such huge savings in computational time of DeePKS with respect to HSE06 also stem from the higher efficiency of NAO basis as compared to the plane-wave basis for large systems. 

\subsubsection{Force and stress}
\begin{figure}[h]
  \centering
  \includegraphics[width=14cm]{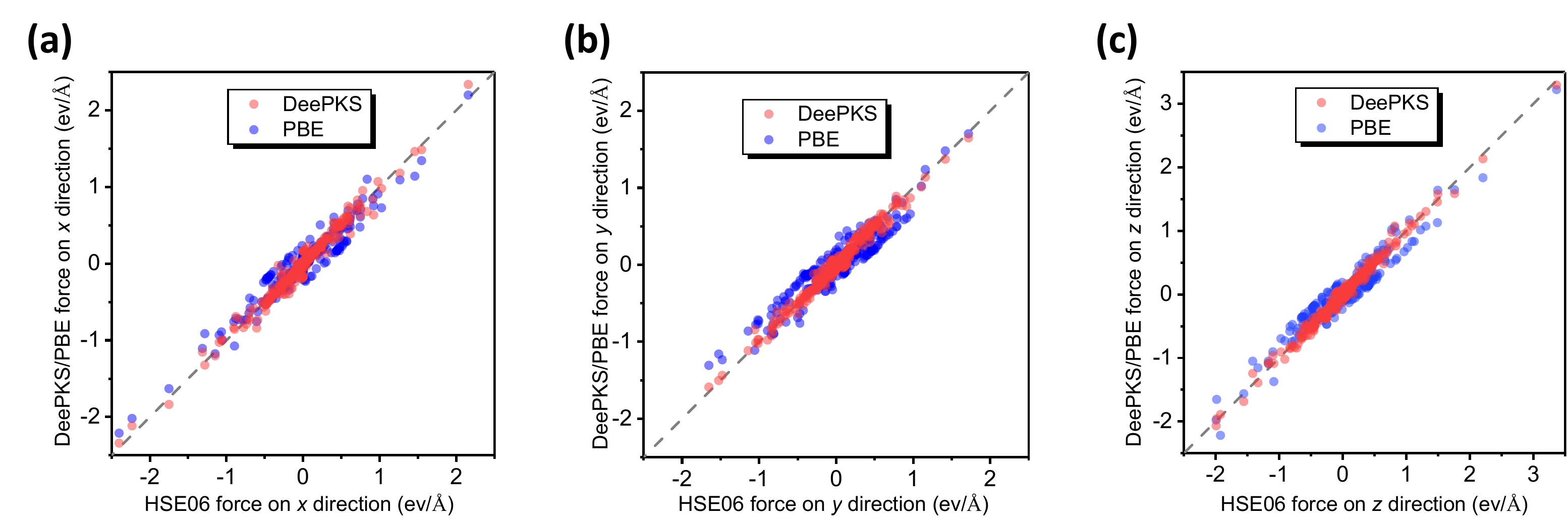}
  \caption{Forces along three directions of hybrid perovskite systems given by DeePKS and PBE with respect to the HSE06 results.}
  \label{fig:force}
\end{figure}

While the absolute energies between DeePKS and HSE06 are not directly comparable owing to the discrepancy between SG15 ONCV pseudopotential applied in DeePKS and PAW method applied in HSE06, we first investigate the accuracy of our DeePKS model in calculating the force of selected systems. Forces are tested on all systems but cubic phase inorganic perovskites, of which the atomic forces remain zero due to their ideal cubic symmetry ($Pm$-3$m$ space group). 
The MAE of 
the force predicted by DeePKS and PBE as compared to the HSE06 result for each 
tested system is listed in Table \ref{tab:force}. Forces along three directions of hybird perovskite systems given by DeePKS and PBE are plotted with respect to the HSE06 values in Figure \ref{fig:force}. It can be seen that while the PBE forces are already close to the HSE06 results, forces calculated by DeePKS are notably more concentrated on the diagonal.
The overall MAE of the force given by 
DeePKS for all tested systems is around 0.030 eV/$\mathring{\textrm{A}}$, while that 
given by PBE is around 0.072 eV/$\mathring{\textrm{A}}$, indicating an improved 
accuracy in the description of the potential energy surface with a sharply higher efficiency as mentioned above. 

\begin{figure}[h]
  \centering
  \includegraphics[width=9cm]{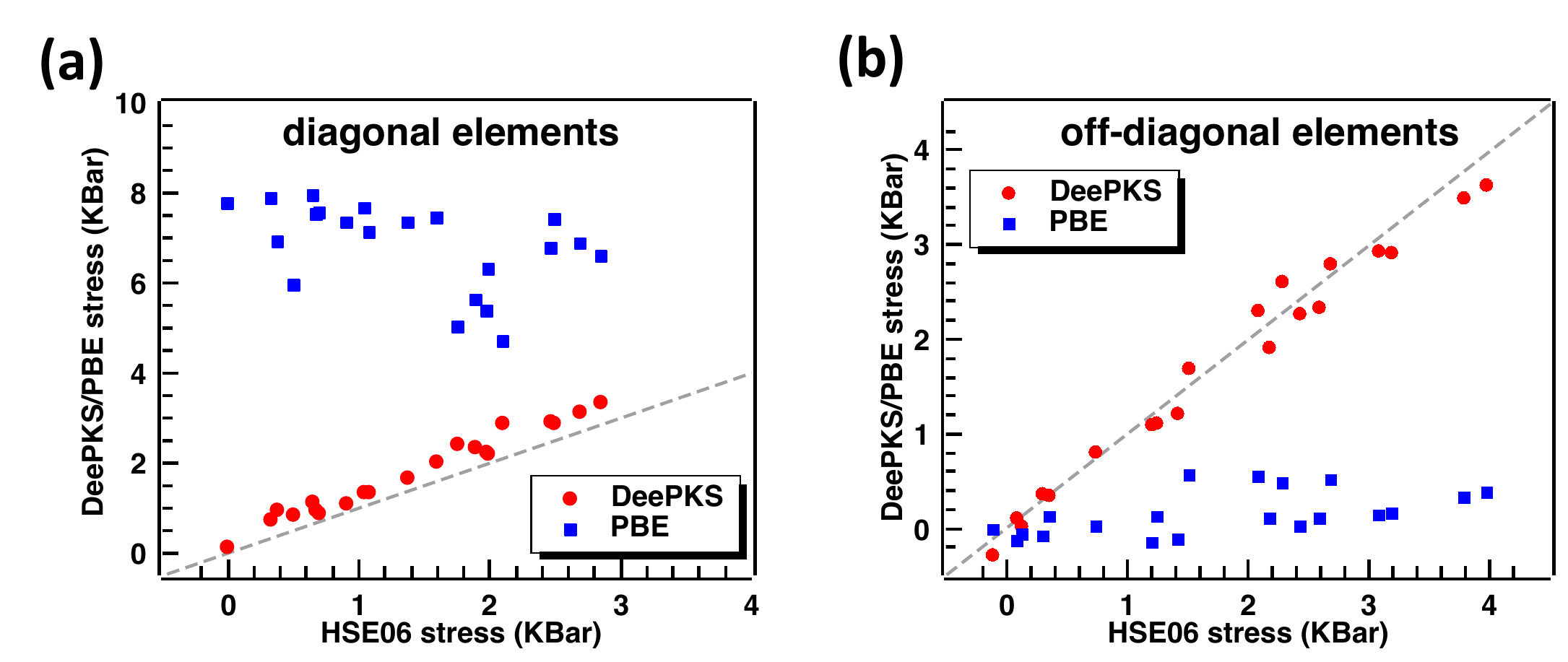}
  \caption{Stresses (diagonal and off-diagonal elements) of hybrid perovskite systems given by DeePKS and PBE with respect to the HSE06 results.}
  \label{fig:stress}
\end{figure}

Compared to the force label, the target stress given by HSE06 is more difficult to be reproduced by DeePKS, owing to the notable deviation between the PBE and HSE06 stresses as mentioned above. Here, we calculate the stress for those hybrid perovskites via DeePKS and PBE, and results are compared with those given by HSE06 as shown in Figure \ref{fig:stress} (while explicit numbers can ben found in Table S5). It can be seen that the diagonal element of the stress calculated by PBE is significantly larger than the HSE06 counterparts, while the off-diagonal elements are severely underestimated. The stress calculated by DeePKS is qualitatively in line with the HSE06 result, with the remarkably reduced MAE for both diagonal and off-diagonal elements of the stress, i.e., 0.394 and 0.168 KBar, respectively (as compared to 5.435 and 1.556 KBar given by PBE). 

\begin{table}[htp]
\centering
\caption{Force MAE for all tested halide perovskites given by DeePKS and PBE with respect to the HSE06 results. All tested non-hybrid organic perovskites are cubic phase, while phases of tested CsPbI$_3$ are indicated by Greek letters. Numbers are shown in the unit of eV/$\mathring{\textrm{A}}$. }
\begin{tabular}{ccc}
\toprule
system &	DeePKS &	PBE	 \\
\midrule
$\beta$-CsPbI$_3$	&	0.0020	&	0.0054	\\
$\gamma$-CsPbI$_3$	&	0.0025	&	0.0053	\\
$\delta$-CsPbI$_3$	&	0.0086	&	0.0268	\\
MAPbCl$_3$	&	0.0245	&	0.0765	\\
MAPbBr$_3$	&	0.0211	&	0.0779	\\
MAPbI$_3$	&	0.0210	&	0.0800	\\
MASnCl$_3$	&	0.0265	&	0.0888	\\
MASnBr$_3$	&	0.0219	&	0.0833	\\
MASnI$_3$	&	0.0317	&	0.0895	\\
FAPbCl$_3$	&	0.0535	&	0.0897	\\
FAPbBr$_3$	&	0.0381	&	0.0887	\\
FAPbI$_3$	&	0.0408	&	0.0927	\\
FASnCl$_3$	&	0.0453	&	0.0897	\\
FASnBr$_3$	&	0.0370	&	0.1053	\\
FASnI$_3$	&	0.0447	&	0.0942	\\
FA$_{0.125}$Cs$_{0.875}$PbI$_3$	&	0.0291	&	0.0445	\\
FA$_{0.25}$Cs$_{0.75}$PbI$_3$	&	0.0281	&	0.0540	\\
FA$_{0.375}$Cs$_{0.625}$PbI$_3$	&	0.0313	&	0.0620	\\
FA$_{0.5}$Cs$_{0.5}$PbI$_3$	&	0.0347	&	0.0782	\\
FA$_{0.625}$Cs$_{0.375}$PbI$_3$	&	0.0383	&	0.0817	\\
FA$_{0.75}$Cs$_{0.25}$PbI$_3$	&	0.0372	&	0.0909	\\
FA$_{0.875}$Cs$_{0.125}$PbI$_3$	&	0.0434	&	0.0905	\\
\midrule
Overall MAE & 0.0301 & 0.0725 \\
\bottomrule
\end{tabular}
\label{tab:force}
\end{table}

\subsubsection{Band gaps and density of states}
\begin{figure}[h]
  \centering
  \includegraphics[width=6cm]{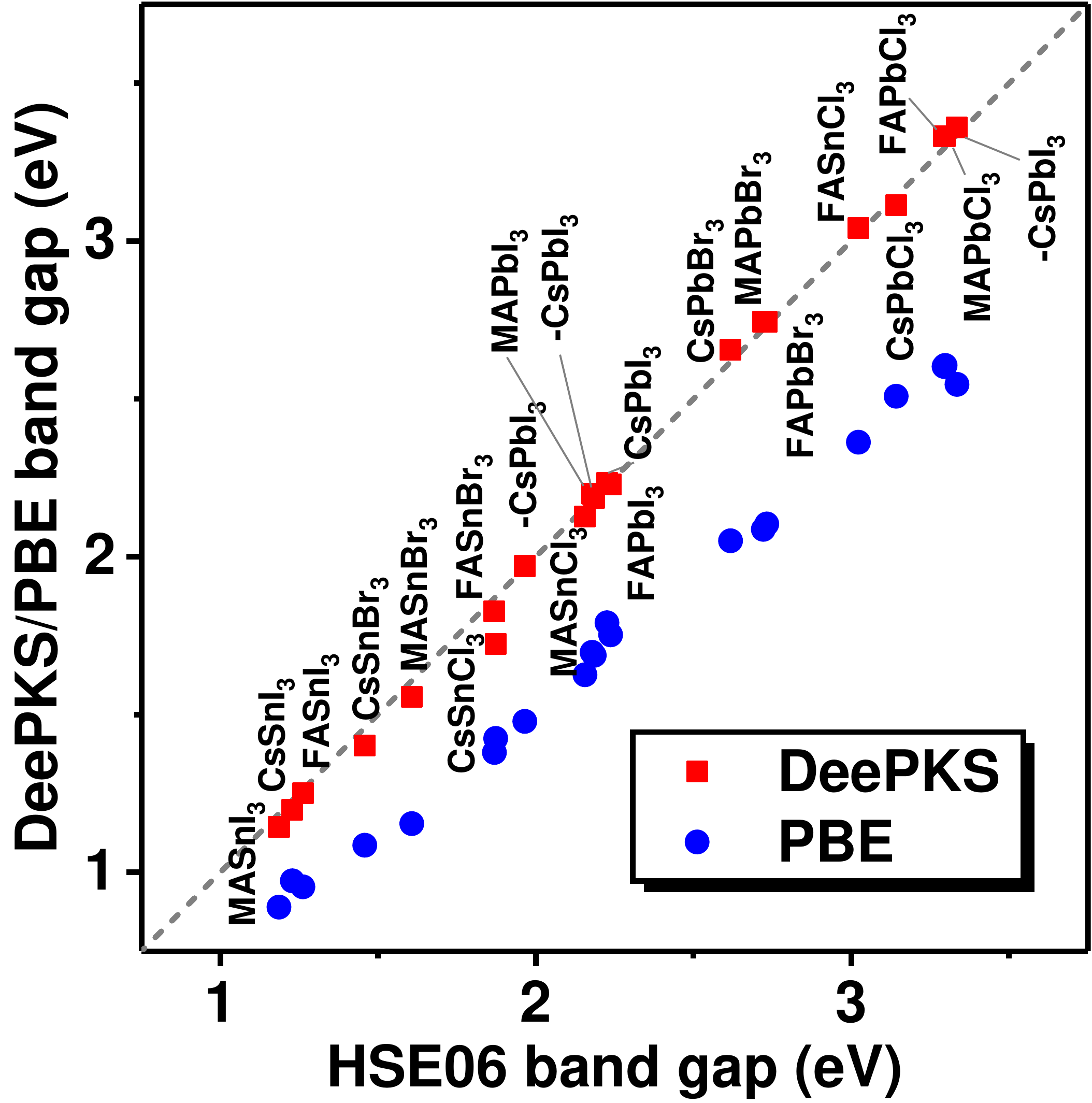}
  \caption{Band gaps predicted by DeePKS and PBE with respect to the HSE06 results for all tested perovskites. All tested non-hybrid perovskites are cubic phase except for those indicated by Greek letters.}
  \label{fig:bandgap}
\end{figure}

Band gaps predicted by DeePKS for all tested systems are compared with the 
results of HSE06 in Table \ref{tab:bandgap}. Values computed via PBE functional are also listed for 
reference.  
As shown in Table \ref{tab:bandgap}, the DeePKS predicted band gaps for all 
tested perovskites (including the hybrid systems and the non-perovskite phase, i.e., $\delta$-CsPbI$_3$) are in excellent agreement with the HSE06 results, with 
an overall MAE 0.0293 eV. The well-known underestimation of the band gap 
values suffered by PBE functional (indicated by the large overall MAE, i.e., 
0.5142 eV) is completely surmounted by the DeePKS model. We visualize the 
band gaps results of DeePKS and PBE with respect to the HSE06 benchmarks in 
Figure \ref{fig:bandgap}. While the PBE predicted values lie significantly lower, the DeePKS results are concentrated on the diagonal line, demonstrating a remarkably 
improved accuracy for the band gap description obtained by the 
DeePKS model. 

\begin{table}[htp]
\centering
\caption{Band gaps for all tested halide perovskites given by HSE06, DeePKS, and PBE. Absolute errors of the DeePKS and PBE results with respect to HSE06 results are listed, as well as the overall mean absolute error (MAE) for all tested systems. All tested non-hybrid perovskites are cubic phase except for those indicated by Greek letters. Numbers are shown in the unit of eV. }
\begin{tabular}{cccccc}
\toprule
system &	HSE06 &	DeePKS	 & DeePKS Abs. Err. &	PBE	 & PBE Abs. Err.\\
\midrule
CsPbCl$_3$	&	3.1423	&	3.1135	&	0.0288	&	2.5079	&	0.6344	\\
CsPbBr$_3$	&	2.6174	&	2.6551	&	0.0377	&	2.0502	&	0.5672	\\
CsPbI$_3$	&	2.2254	&	2.2331	&	0.0077	&	1.7906	&	0.4348	\\
$\beta$-CsPbI$_3$	&	1.9650	&	1.9703	&	0.0053	&	1.4782	&	0.4868	\\
$\gamma$-CsPbI$_3$	&	2.1854	&	2.1861	&	0.0007	&	1.6873	&	0.4981	\\
$\delta$-CsPbI$_3$	&	3.3353	&	3.3603	&	0.0250	&	2.5454	&	0.7899	\\
CsSnCl$_3$	&	1.8727	&	1.7226	&	0.1501	&	1.4237	&	0.4490	\\
CsSnBr$_3$	&	1.4584	&	1.3750	&	0.0834	&	1.0856	&	0.3728	\\
CsSnI$_3$	&	1.2288	&	1.1979	&	0.0309	&	0.9730	&	0.2558	\\
MAPbCl$_3$	&	3.2972	&	3.3318	&	0.0346	&	2.6055	&	0.6917	\\
MAPbBr$_3$	&	2.7211	&	2.7439	&	0.0228	&	2.0861	&	0.6350	\\
MAPbI$_3$	&	2.1789	&	2.1981	&	0.0192	&	1.6966	&	0.4823	\\
MASnCl$_3$	&	2.1558	&	2.1264	&	0.0294	&	1.6259	&	0.5299	\\
MASnBr$_3$	&	1.6072	&	1.5557	&	0.0515	&	1.1542	&	0.4530	\\
MASnI$_3$	&	1.1858	&	1.1443	&	0.0415	&	0.8897	&	0.2961	\\
FAPbCl$_3$	&	3.2938	&	3.3319	&	0.0381	&	2.6024	&	0.6914	\\
FAPbBr$_3$	&	2.7325	&	2.7443	&	0.0118	&	2.1036	&	0.6289	\\
FAPbI$_3$	&	2.2374	&	2.2284	&	0.0090	&	1.7521	&	0.4853	\\
FASnCl$_3$	&	3.0226	&	3.0412	&	0.0186	&	2.3624	&	0.6602	\\
FASnBr$_3$	&	1.8691	&	1.8273	&	0.0418	&	1.3795	&	0.4896	\\
FASnI$_3$	&	1.2622	&	1.2502	&	0.0120	&	0.9536	&	0.3086	\\
FA$_{0.125}$Cs$_{0.875}$PbI$_3$	&	2.3968	&	2.3805	&	0.0163	&	1.8556	&	0.5412	\\
FA$_{0.25}$Cs$_{0.75}$PbI$_3$	&	2.3929	&	2.3952	&	0.0023	&	1.8499	&	0.5430	\\
FA$_{0.375}$Cs$_{0.625}$PbI$_3$	&	2.0071	&	1.9875	&	0.0196	&	1.5313	&	0.4758	\\
FA$_{0.5}$Cs$_{0.5}$PbI$_3$	&	2.3547	&	2.3428	&	0.0119	&	1.8072	&	0.5475	\\
FA$_{0.625}$Cs$_{0.375}$PbI$_3$	&	2.0222	&	2.0061	&	0.0161	&	1.5499	&	0.4723	\\
FA$_{0.75}$Cs$_{0.25}$PbI$_3$	&	2.0145	&	1.9997	&	0.0148	&	1.5411	&	0.4734	\\
FA$_{0.875}$Cs$_{0.125}$PbI$_3$	&	2.0808	&	2.0429	&	0.0379	&	1.5760	&	0.5048	\\
\midrule
\multicolumn{3}{c}{Overall MAE} & 0.0293 & & 0.5142 \\
\bottomrule
\end{tabular}
\label{tab:bandgap}
\end{table}

\begin{figure}[h]
  \centering
  \includegraphics[width=14cm]{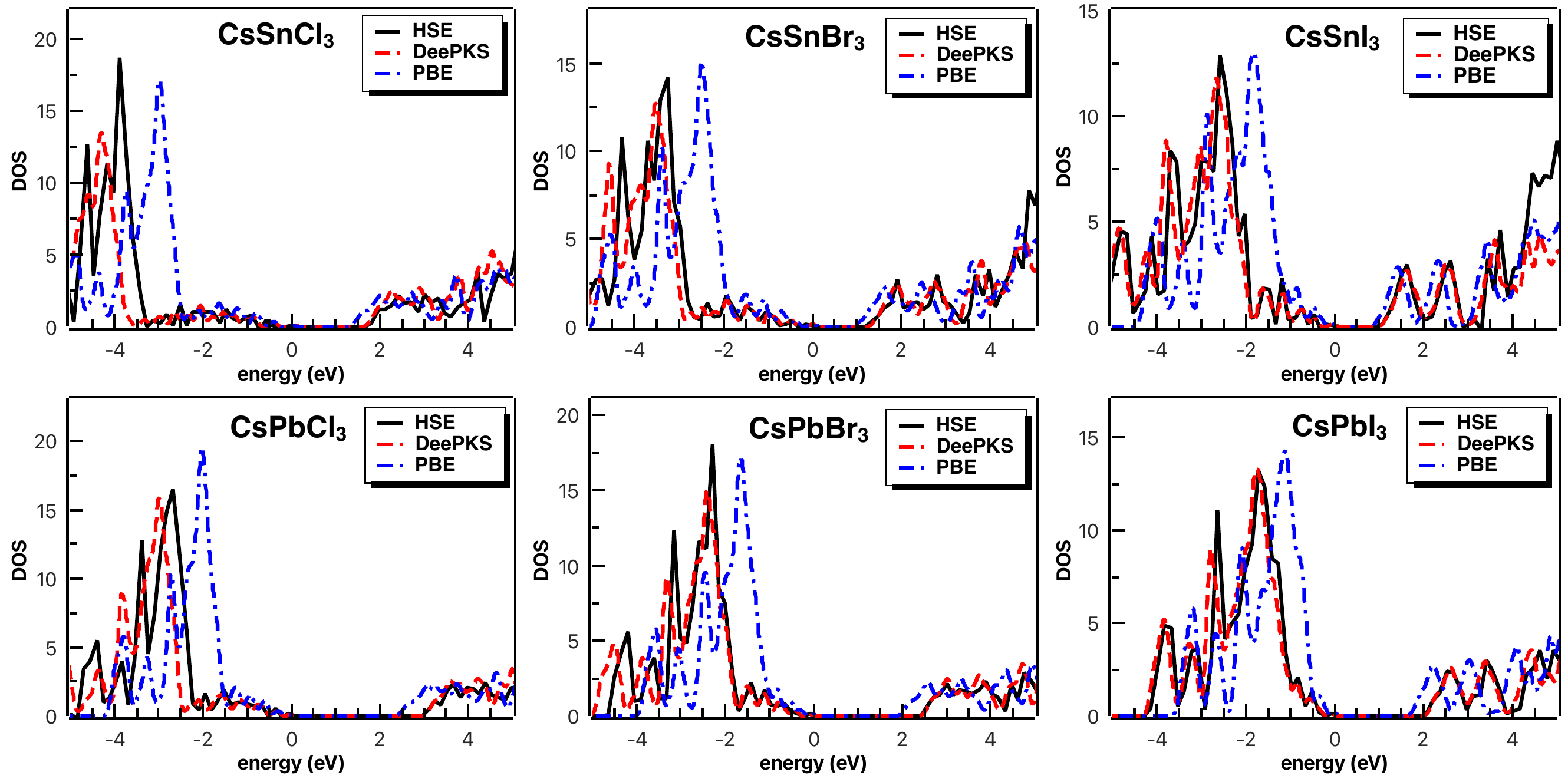}
  \caption{Density of states (DOS) for cubic phase inorganic halide perovskites, i.e., CsBX$_3$ (B=Sn, Pb; X=Cl, Br, I).}
  \label{fig:csbx3_dos}
\end{figure}

\begin{figure}[h]
  \centering
  \includegraphics[width=14cm]{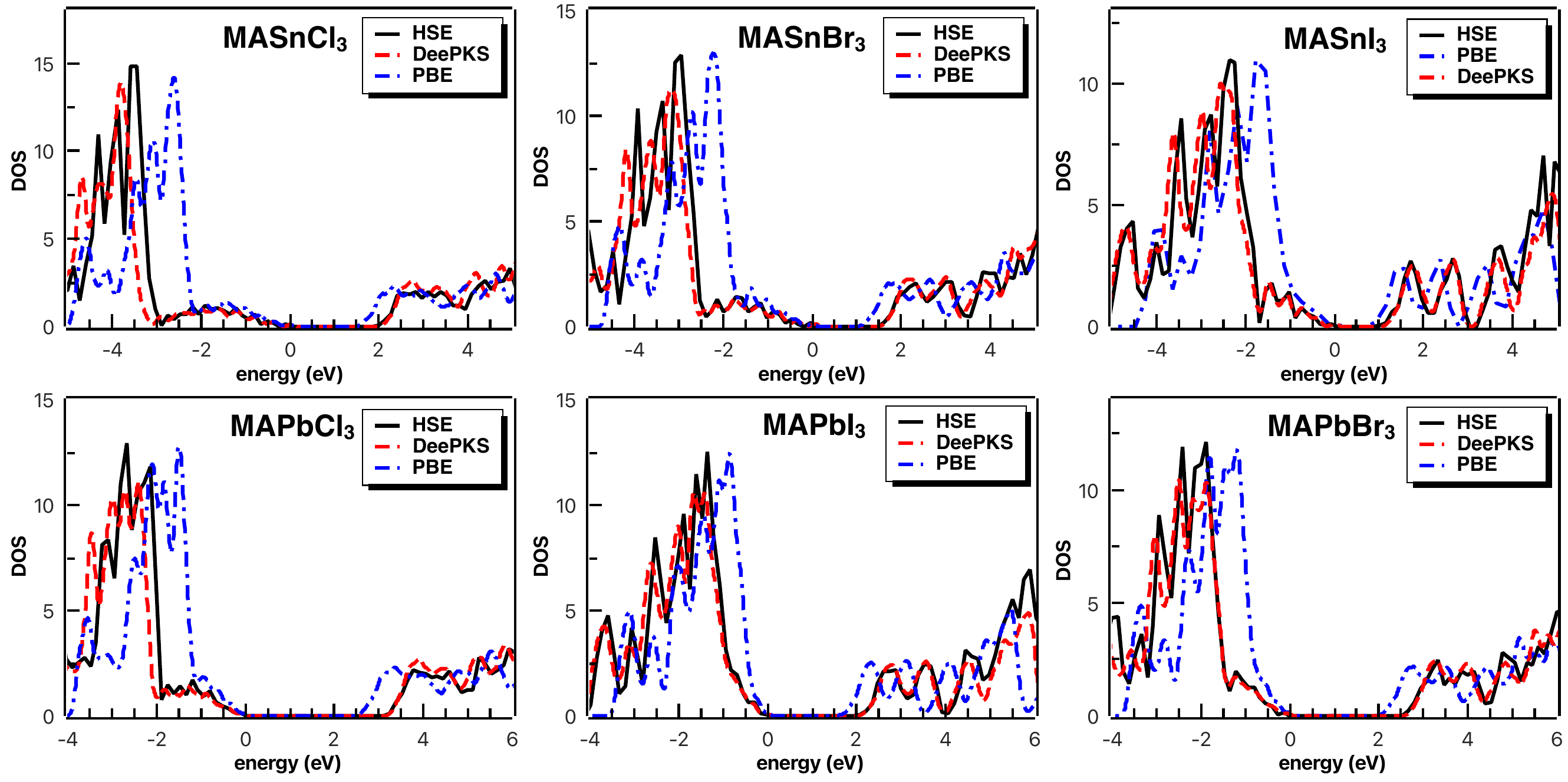}
  \caption{DOS for cubic phase organic halide perovskites, i.e., MABX$_3$ (B=Sn, Pb; X=Cl, Br, I).}
  \label{fig:mabx3_dos}
\end{figure}

\begin{figure}[h]
  \centering
  \includegraphics[width=14cm]{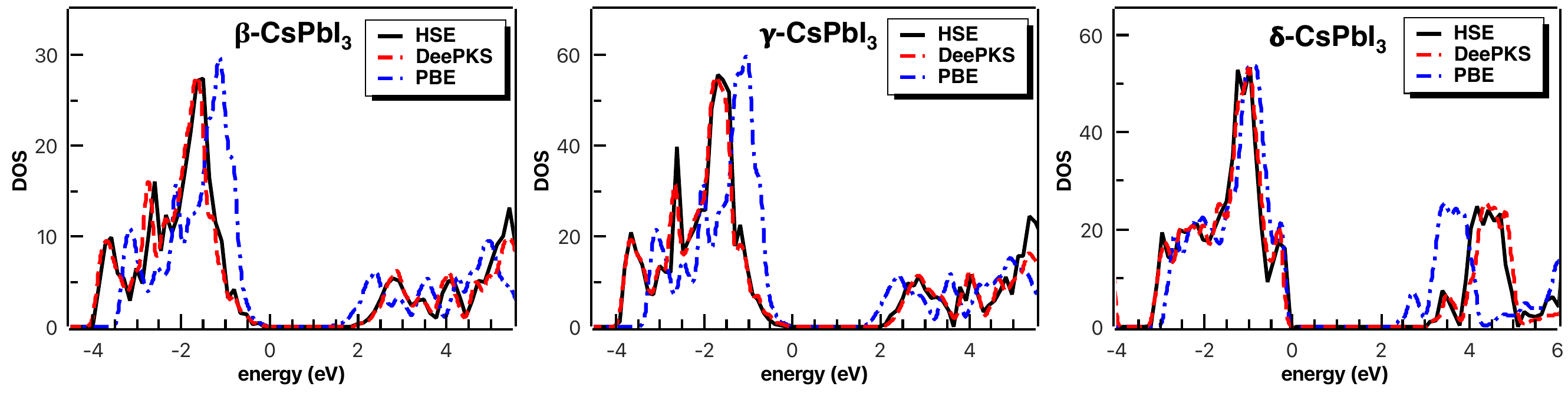}
  \caption{DOS for non-cubic phase CsPbI$_3$.}
  \label{fig:non_cubic}
\end{figure}

\begin{figure}[h]
  \centering
  \includegraphics[width=14cm]{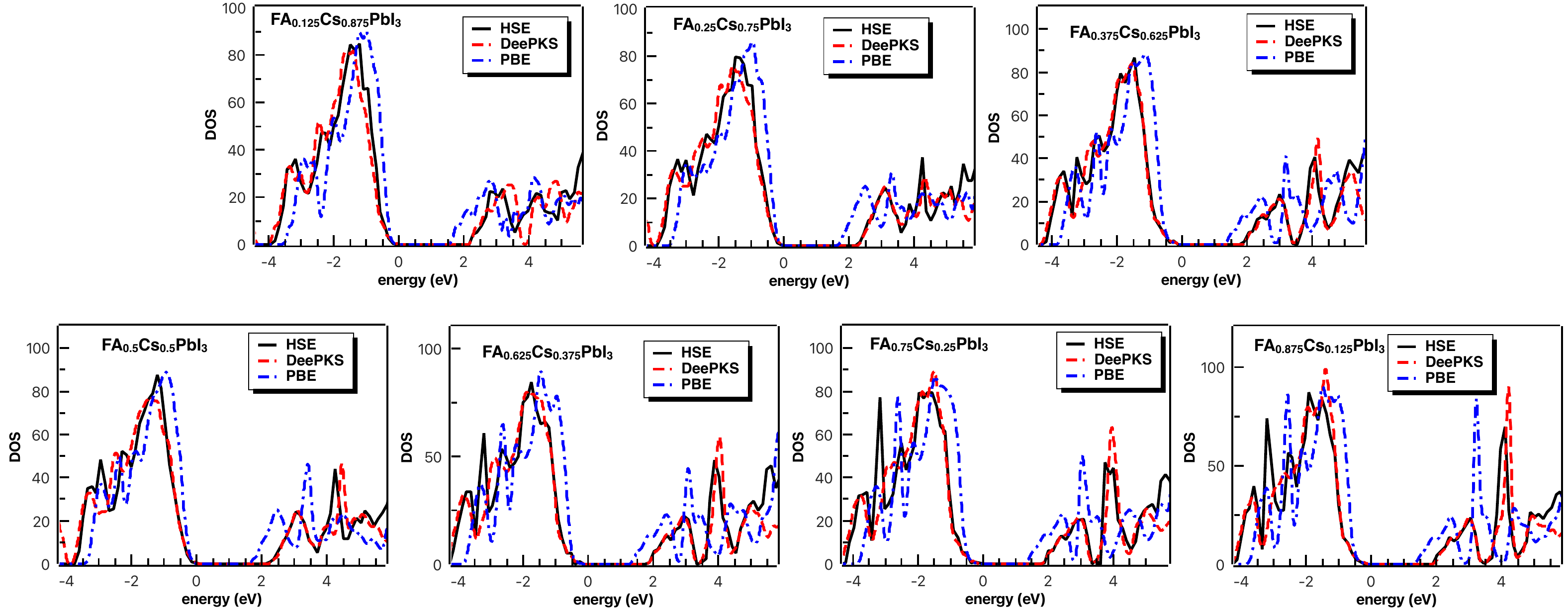}
  \caption{DOS for hybrid organic-inorganic halide perovskites, i.e., FA$_a$Cs$_{1.0-a}$PbI$_3$.}
  \label{fig:hybrid_dos}
\end{figure}

Next, we examine the DOS of various halide perovskites described by DeePKS.
Even though the DOS does not directly serve as one of the training labels 
to be fitted by the neural network, we can see from Figures \ref{fig:csbx3_dos} to \ref{fig:hybrid_dos} that for all tested 
halide perovskites, the overall DOSs near the Fermi energy 
predicted by DeePKS precisely match with those given by HSE06, while those of PBE 
exhibit significantly smaller band gaps. Such accurate description 
of the DOS given by DeePKS lies in the fact that bands predicted by 
PBE near the Fermi energy possess an overall reasonable structure 
(as speculated by the similar shape of the DOS predicted by HSE06 
and PBE) but with a much narrower gap. By fitting the band gap for each $k$ point, such deficiency of PBE can be largely mitigated. 

\subsubsection{Spin-orbit coupling effect on Pb-containing systems}
\begin{figure}[h]
  \centering
  \includegraphics[width=14cm]{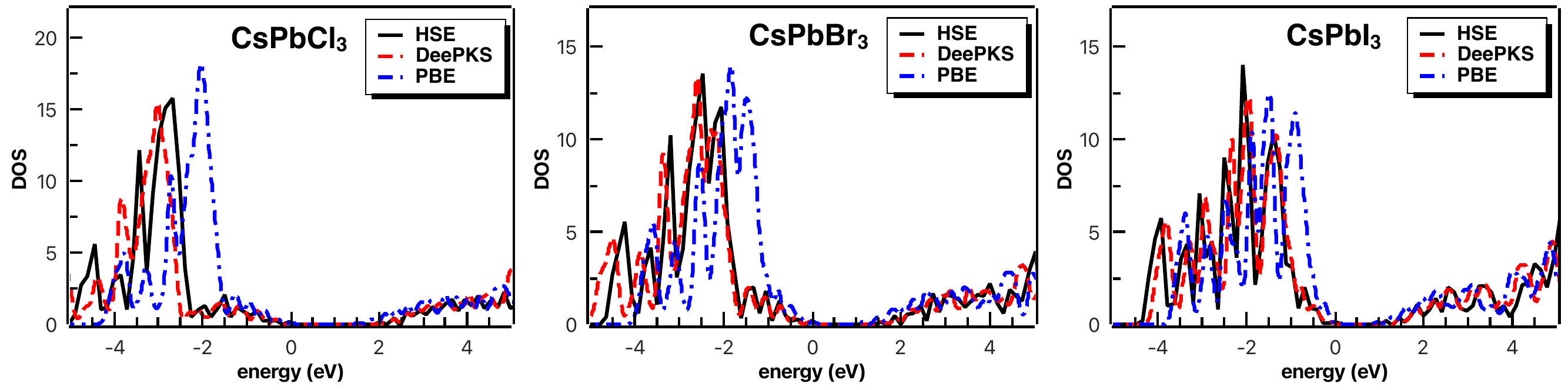}
  \caption{DOS for Pb-containing cubic phase inorganic halide perovskites, i.e., CsPbX$_3$ (X=Cl, Br, I) with the SOC effect taken into account.}
  \label{fig:soc_dos}
\end{figure}

It is well-known that the SOC effect plays an 
important role to the magnitude of the band gaps in perovskite systems,\cite{doi:10.1021/jz401532q,doi:10.1021/acs.jpclett.0c02135} 
especially for Pb-containing perovskite.\cite{lead_soc_nc,doi:10.1021/acs.jpcc.1c09594} The inclusion of the SOC effect may 
significantly narrow the band gap for APbX$_3$ and thus becomes essential 
for accurate property prediction in practice.\cite{doi:10.1021/acs.jpcc.1c09594,C9TA12263J} Taking CsPbX$_3$ as examples, 
we investigate the performance of DeePKS model with the inclusion of the SOC 
effect. To do that, the full relativistic (FR) version of the SG15 ONCV pseudopotentials\cite{doi:10.1021/acs.jctc.6b00114} is applied for all involved elements.
As shown in Table \ref{tab:bandgap_soc} and Figure \ref{fig:soc_dos}, while the 
SOC effect has not been taken into account during our iterative DeePKS 
training process, the DeePKS+SOC offers closely matched band gap values as well as 
the DOS for three tested systems as compared to HSE06+SOC. Indeed, the relativistic effect is mainly handled 
in the pseudopotential part, either via the PAW method (for HSE06 functional performed in VASP) or via the norm-conserving pseudopotentials (for DeePKS model performed in ABACUS). Even 
though the scalar relativistic approximation is applied in PAW for the valance electron\cite{https://doi.org/10.1002/jcc.21057} while the FR 
effect is employed in ONCV pseudopotentials, the influence of such 
discrepancy is subtle for our tested systems, as demonstrated by the almost 
identical DOS given by PBE+SOC calculations performed in VASP with PAW 
and that in ABACUS with NAO and the FR version of ONCV pseudopotentials (Figure S1). 
Therefore, the HSE06+SOC results can be successfully reproduced by DeePKS+SOC as long as the DeePKS model provides a close enough description of the electronic structure as compared to the HSE06 functional, which has been extensively evinced in former sections. 

\begin{table}[htp]
\centering
\caption{Band gaps for cubic phase CsPbX$_3$ given by HSE06, DeePKS, and PBE with the SOC effect taken into account. Absolute errors of the DeePKS and PBE results with respect to HSE06 results are listed. Numbers are shown in the unit of eV. }
\begin{tabular}{cccccc}
\toprule
system &	HSE06 &	DeePKS	 & DeePKS Abs. Err. &	PBE	 & PBE Abs. Err.\\
\midrule
CsPbCl$_3$	&	2.1187	&	2.0777	&	0.0410	&	1.5382	&	0.5805	\\
CsPbBr$_3$	&	1.6298	&	1.6295	&	0.0003	&	1.1261	&	0.5037	\\
CsPbI$_3$	&	1.2333	&	1.2456	&	0.0123	&	0.9353	&	0.2980	\\
\bottomrule
\end{tabular}
\label{tab:bandgap_soc}
\end{table}

\subsubsection{Equation of states for cubic inorganic perovskites}
\begin{figure}[h]
  \centering
  \includegraphics[width=14cm]{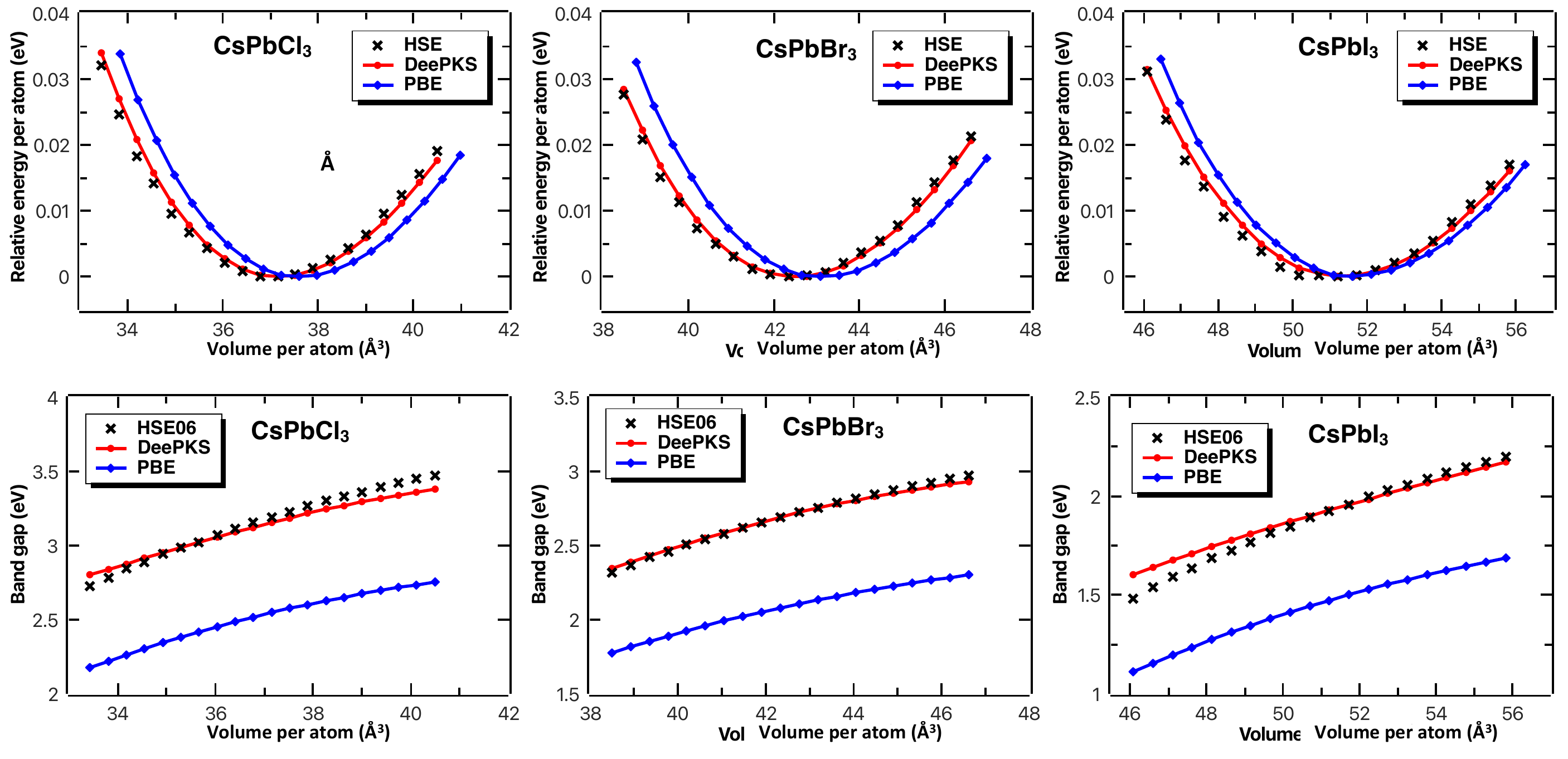}
  \caption{Equation of states (EoS) (upper panel) and band gap values (lower panel) with respect to the size of the unit cell of CsPbBr$_3$, CsPbCl$_3$, CsPbBr$_3$, and CsPbI$_3$ computed with HSE06, DeePKS, and PBE.}
  \label{fig:eos}
\end{figure}

Finally, we examine the capability of our DeePKS model in predicting mechanical properties by performing the EoS calculation for three cubic phase inorganic perovskites, i.e., CsPbCl$_3$, CsPbBr$_3$, and CsPbI$_3$. A cell relaxation is first performed with DeePKS model and PBE functional (with the same input settings as those applied in former single-point calculation). Resulting optimized volumes are then scaled from 0.9 to 1.1 with a step of 0.01 to compute the total energy at different volumes. Due to the extremely high computational cost of HSE06 functional, we skip the relaxation step by applying the DeePKS-optmized structure at each volume and perform a single point SCF calculation via HSE06 funcitonal. Since the absolute energies given by ONCV pseudopotentials and PAW method are not directly comparable, we plot the relative energies with respect to the lowest energy for each functional. As shown in the upper panel of Figure \ref{fig:eos}, the DeePKS computed EoS for these systems is in line with the HSE06 counterpart, while PBE predicts a larger volume of the equilibrium structure. Change of the band gap values with respect to the size of the unit cell is also explored as shown in the lower panel of Figure \ref{fig:eos}. Closely matched band gaps are obtained via DeePKS and HSE06, indicating that the band gap tuning by adjusting the pressure can be accurately predicted via DeePKS model. 

\section{Conclusion and outlook}
In this work, with 460 configurations over seven types of halide perovskites, we have established a general DeePKS model that is applicable to a variety of combinations, i.e., ABX$_3$ (A=FA, MA, Cs; B=Sn, Pb; X=Cl, Br, I) as well as the organic-inorganic hybrid perovskites, with the accuracy of HSE06 in predicting electronic structure properties such as force, stress, band gap, and DOS, and the efficiency comparable to that of PBE. For all tested systems, especially hybrid perovskites with large unit cells, DeePKS offers orders of magnitude savings in computational cost as compared to HSE06. While the MAEs of the energy and force predicted by DeePKS with respect to HSE06 are reduced by a factor of two as compared to those of PBE, the MAEs for the stress and band gap are reduced at least one order of magnitude. Specifically, the MAE for the band gap given by DeePKS over 28 tested systems is less than 0.03 eV, demonstrating a promising alleviation of the underestimation issue suffered by PBE, of which the MAE is larger than 0.5 eV. The DOS predicted by DeePKS also closely matches with the target method in the vicinity of the Fermi energy. 

For Pb-containing perovskites, the SOC effect needs to be taken into account so as to offer a quantitatively correct description for the band gap and DOS. We have shown that DeePKS+SOC is able to offer highly consistent band gap and DOS as compared to HSE06+SOC for Pb-containing systems even though the SOC effect is not considered during the iterative training process. In addition, mechanical properties given by DeePKS have also been validated for inorganic halide perovskites by performing the EoS calculation, with highly consistent band gaps with respect to the size of the unit cell as compared to HSE06. In summary, with the closely-matched electronic structure description with respect to the HSE06 functional and remarkbaly reduced computational cost, our DeePKS model can be extensively applied to precisely predict a variety of properties and/or generate highly accurate labels for machine learning based potential energy models for the family of halide perovskites, facilitating the discovery and design of better photovoltaics based on such systems.

\begin{acknowledgement}
The authors thank Dr. Han Wang and Dr. Mohan Chen for their helpful conversations and computational guidance. The authors gratefully acknowledge the funding support from the AI for Science Institute, Beijing (AISI) and DP Technology Corporation. The computing resource of this work was provided by the Bohrium Cloud Platform (https://bohrium.dp.tech), which is supported by DP Technology. The work of Y.C. was supported by the Computational Chemical Sciences Center: Chemistry in Solution and at Interfaces (CSI) funded by DOE Award DE-SC0019394 and a gift from iFlytek to Princeton University.
\end{acknowledgement}

\begin{suppinfo}
Details of the SCF calculation (including input parameters and run time statistics) and the DeePKS iterative training procedure. MAEs of various electronic structure properties for the training and test sets given by the DeePKS model.   
\end{suppinfo}

\bibliography{main}

\providecommand{\latin}[1]{#1}
\makeatletter
\providecommand{\doi}
  {\begingroup\let\do\@makeother\dospecials
  \catcode`\{=1 \catcode`\}=2 \doi@aux}
\providecommand{\doi@aux}[1]{\endgroup\texttt{#1}}
\makeatother
\providecommand*\mcitethebibliography{\thebibliography}
\csname @ifundefined\endcsname{endmcitethebibliography}
  {\let\endmcitethebibliography\endthebibliography}{}
\begin{mcitethebibliography}{58}
\providecommand*\natexlab[1]{#1}
\providecommand*\mciteSetBstSublistMode[1]{}
\providecommand*\mciteSetBstMaxWidthForm[2]{}
\providecommand*\mciteBstWouldAddEndPuncttrue
  {\def\EndOfBibitem{\unskip.}}
\providecommand*\mciteBstWouldAddEndPunctfalse
  {\let\EndOfBibitem\relax}
\providecommand*\mciteSetBstMidEndSepPunct[3]{}
\providecommand*\mciteSetBstSublistLabelBeginEnd[3]{}
\providecommand*\EndOfBibitem{}
\mciteSetBstSublistMode{f}
\mciteSetBstMaxWidthForm{subitem}{(\alph{mcitesubitemcount})}
\mciteSetBstSublistLabelBeginEnd
  {\mcitemaxwidthsubitemform\space}
  {\relax}
  {\relax}

\bibitem[Chu and Majumdar(2012)Chu, and Majumdar]{energy1}
Chu,~S.; Majumdar,~A. Opportunities and Challenges for a Sustainable Energy
  Future. \emph{Nature} \textbf{2012}, \emph{488}, 294--303\relax
\mciteBstWouldAddEndPuncttrue
\mciteSetBstMidEndSepPunct{\mcitedefaultmidpunct}
{\mcitedefaultendpunct}{\mcitedefaultseppunct}\relax
\EndOfBibitem
\bibitem[Deng \latin{et~al.}(2014)Deng, Xu, Liu, and Mancl]{energy2}
Deng,~Y.; Xu,~J.; Liu,~Y.; Mancl,~K. Biogas as a Sustainable Energy Source in
  China: Regional Development Strategy Application and Decision Making.
  \emph{Renewable Sustainable Energy Rev.} \textbf{2014}, \emph{35},
  294--303\relax
\mciteBstWouldAddEndPuncttrue
\mciteSetBstMidEndSepPunct{\mcitedefaultmidpunct}
{\mcitedefaultendpunct}{\mcitedefaultseppunct}\relax
\EndOfBibitem
\bibitem[Chu \latin{et~al.}(2016)Chu, Cui, and Liu]{energy3}
Chu,~S.; Cui,~Y.; Liu,~N. The Path towards Sustainable Energy. \emph{Nat.
  Mater.} \textbf{2016}, \emph{16}, 16--22\relax
\mciteBstWouldAddEndPuncttrue
\mciteSetBstMidEndSepPunct{\mcitedefaultmidpunct}
{\mcitedefaultendpunct}{\mcitedefaultseppunct}\relax
\EndOfBibitem
\bibitem[Parida \latin{et~al.}(2011)Parida, Iniyan, and Goic]{energy4}
Parida,~B.; Iniyan,~S.; Goic,~R. A Review of Solar Photovoltaic Technologies.
  \emph{Renewable Sustainable Energy Rev.} \textbf{2011}, \emph{15},
  1625--1636\relax
\mciteBstWouldAddEndPuncttrue
\mciteSetBstMidEndSepPunct{\mcitedefaultmidpunct}
{\mcitedefaultendpunct}{\mcitedefaultseppunct}\relax
\EndOfBibitem
\bibitem[Riede \latin{et~al.}(2021)Riede, Spoltore, and Leo]{energy5}
Riede,~M.; Spoltore,~D.; Leo,~K. Organic Solar Cells—The Path to Commercial
  Success. \emph{Adv. Energy Mater.} \textbf{2021}, \emph{11}, 2002653\relax
\mciteBstWouldAddEndPuncttrue
\mciteSetBstMidEndSepPunct{\mcitedefaultmidpunct}
{\mcitedefaultendpunct}{\mcitedefaultseppunct}\relax
\EndOfBibitem
\bibitem[Nayak \latin{et~al.}(2019)Nayak, Mahesh, Snaith, and Cahen]{pv1}
Nayak,~P.~K.; Mahesh,~S.; Snaith,~H.~J.; Cahen,~D. Photovoltaic Solar Cell
  Technologies: Analysing the State of the Art. \emph{Nat. Rev. Mater.}
  \textbf{2019}, \emph{4}, 269--285\relax
\mciteBstWouldAddEndPuncttrue
\mciteSetBstMidEndSepPunct{\mcitedefaultmidpunct}
{\mcitedefaultendpunct}{\mcitedefaultseppunct}\relax
\EndOfBibitem
\bibitem[Gong \latin{et~al.}(2012)Gong, Liang, and Sumathy]{pv2}
Gong,~J.; Liang,~J.; Sumathy,~K. Review on dye-sensitized solar cells (DSSCs):
  Fundamental concepts and novel materials. \emph{Renewable Sustainable Energy
  Rev.} \textbf{2012}, \emph{16}, 5848--5860\relax
\mciteBstWouldAddEndPuncttrue
\mciteSetBstMidEndSepPunct{\mcitedefaultmidpunct}
{\mcitedefaultendpunct}{\mcitedefaultseppunct}\relax
\EndOfBibitem
\bibitem[Shah \latin{et~al.}(1999)Shah, Torres, Tscharner, Wyrsch, and
  Keppner]{pv3}
Shah,~A.; Torres,~P.; Tscharner,~R.; Wyrsch,~N.; Keppner,~H. Photovoltaic
  Technology: The Case for Thin-Film Solar Cells. \emph{Science} \textbf{1999},
  \emph{285}, 692--698\relax
\mciteBstWouldAddEndPuncttrue
\mciteSetBstMidEndSepPunct{\mcitedefaultmidpunct}
{\mcitedefaultendpunct}{\mcitedefaultseppunct}\relax
\EndOfBibitem
\bibitem[Polman \latin{et~al.}(2016)Polman, Knight, Garnett, Ehrler, and
  Sinke]{pv4}
Polman,~A.; Knight,~M.; Garnett,~E.~C.; Ehrler,~B.; Sinke,~W.~C. Photovoltaic
  materials: Present efficiencies and future challenges. \emph{Science}
  \textbf{2016}, \emph{352}, aad4424\relax
\mciteBstWouldAddEndPuncttrue
\mciteSetBstMidEndSepPunct{\mcitedefaultmidpunct}
{\mcitedefaultendpunct}{\mcitedefaultseppunct}\relax
\EndOfBibitem
\bibitem[Andreani \latin{et~al.}(2019)Andreani, Bozzola, Kowalczewski,
  Liscidini, and Redorici]{silicon1}
Andreani,~L.~C.; Bozzola,~A.; Kowalczewski,~P.; Liscidini,~M.; Redorici,~L.
  Silicon solar cells: toward the efficiency limits. \emph{Adv. Phys.: X}
  \textbf{2019}, \emph{4}, 1548305\relax
\mciteBstWouldAddEndPuncttrue
\mciteSetBstMidEndSepPunct{\mcitedefaultmidpunct}
{\mcitedefaultendpunct}{\mcitedefaultseppunct}\relax
\EndOfBibitem
\bibitem[Sun \latin{et~al.}(2022)Sun, Chen, He, Li, Wang, Yan, and
  Zhang]{silicon2}
Sun,~Z.; Chen,~X.; He,~Y.; Li,~J.; Wang,~J.; Yan,~H.; Zhang,~Y. Toward
  Efficiency Limits of Crystalline Silicon Solar Cells: Recent Progress in
  High-Efficiency Silicon Heterojunction Solar Cells. \emph{Adv. Energy Mater.}
  \textbf{2022}, \emph{12}, 2200015\relax
\mciteBstWouldAddEndPuncttrue
\mciteSetBstMidEndSepPunct{\mcitedefaultmidpunct}
{\mcitedefaultendpunct}{\mcitedefaultseppunct}\relax
\EndOfBibitem
\bibitem[Battaglia \latin{et~al.}(2016)Battaglia, Cuevas, and
  De~Wolf]{silicon3}
Battaglia,~C.; Cuevas,~A.; De~Wolf,~S. High-efficiency crystalline silicon
  solar cells: status and perspectives. \emph{Energy Environ. Sci.}
  \textbf{2016}, \emph{9}, 1552--1576\relax
\mciteBstWouldAddEndPuncttrue
\mciteSetBstMidEndSepPunct{\mcitedefaultmidpunct}
{\mcitedefaultendpunct}{\mcitedefaultseppunct}\relax
\EndOfBibitem
\bibitem[Peng and Lee(2011)Peng, and Lee]{silicon4}
Peng,~K.-Q.; Lee,~S.-T. Silicon Nanowires for Photovoltaic Solar Energy
  Conversion. \emph{Adv. Mater.} \textbf{2011}, \emph{23}, 198--215\relax
\mciteBstWouldAddEndPuncttrue
\mciteSetBstMidEndSepPunct{\mcitedefaultmidpunct}
{\mcitedefaultendpunct}{\mcitedefaultseppunct}\relax
\EndOfBibitem
\bibitem[Jin \latin{et~al.}(2023)Jin, Wang, Ma, Shen, Belfiore, Bao, and
  Tang]{opv1}
Jin,~J.; Wang,~Q.; Ma,~K.; Shen,~W.; Belfiore,~L.~A.; Bao,~X.; Tang,~J. Recent
  Developments of Polymer Solar Cells with Photovoltaic Performance over 17\%.
  \emph{Adv. Funct. Mater.} \textbf{2023}, \emph{33}, 2213324\relax
\mciteBstWouldAddEndPuncttrue
\mciteSetBstMidEndSepPunct{\mcitedefaultmidpunct}
{\mcitedefaultendpunct}{\mcitedefaultseppunct}\relax
\EndOfBibitem
\bibitem[Wang \latin{et~al.}(2022)Wang, Zhao, Kan, Xie, and Pan]{dssc1}
Wang,~X.; Zhao,~B.; Kan,~W.; Xie,~Y.; Pan,~K. Review on Low-Cost Counter
  Electrode Materials for Dye-Sensitized Solar Cells: Effective Strategy to
  Improve Photovoltaic Performance. \emph{Adv. Mater. Interfaces}
  \textbf{2022}, \emph{9}, 2101229\relax
\mciteBstWouldAddEndPuncttrue
\mciteSetBstMidEndSepPunct{\mcitedefaultmidpunct}
{\mcitedefaultendpunct}{\mcitedefaultseppunct}\relax
\EndOfBibitem
\bibitem[Datt \latin{et~al.}(2022)Datt, Bishnoi, Lee, Arya, Gupta, Gupta, and
  Tsoi]{opv2}
Datt,~R.; Bishnoi,~S.; Lee,~H. K.~H.; Arya,~S.; Gupta,~S.; Gupta,~V.;
  Tsoi,~W.~C. Down-conversion materials for organic solar cells: Progress,
  challenges, and perspectives. \emph{Aggregate} \textbf{2022}, \emph{3},
  e185\relax
\mciteBstWouldAddEndPuncttrue
\mciteSetBstMidEndSepPunct{\mcitedefaultmidpunct}
{\mcitedefaultendpunct}{\mcitedefaultseppunct}\relax
\EndOfBibitem
\bibitem[Zhang \latin{et~al.}(2022)Zhang, Lin, Qi, Heumüller, Distler,
  Egelhaaf, Li, Chow, Brabec, Jen, and Yip]{opv3}
Zhang,~G.; Lin,~F.~R.; Qi,~F.; Heumüller,~T.; Distler,~A.; Egelhaaf,~H.-J.;
  Li,~N.; Chow,~P. C.~Y.; Brabec,~C.~J.; Jen,~A. K.-Y.; Yip,~H.-L. Renewed
  Prospects for Organic Photovoltaics. \emph{Chem. Rev.} \textbf{2022},
  \emph{122}, 14180--14274, PMID: 35929847\relax
\mciteBstWouldAddEndPuncttrue
\mciteSetBstMidEndSepPunct{\mcitedefaultmidpunct}
{\mcitedefaultendpunct}{\mcitedefaultseppunct}\relax
\EndOfBibitem
\bibitem[Lin and Zhan(2016)Lin, and Zhan]{opv_pce1}
Lin,~Y.; Zhan,~X. Oligomer Molecules for Efficient Organic Photovoltaics.
  \emph{Acc. Chem. Res.} \textbf{2016}, \emph{49}, 175--183\relax
\mciteBstWouldAddEndPuncttrue
\mciteSetBstMidEndSepPunct{\mcitedefaultmidpunct}
{\mcitedefaultendpunct}{\mcitedefaultseppunct}\relax
\EndOfBibitem
\bibitem[Bernardo \latin{et~al.}(2021)Bernardo, Lopes, Lidzey, and
  Mendes]{opv_pce2}
Bernardo,~G.; Lopes,~T.; Lidzey,~D.~G.; Mendes,~A. Progress in Upscaling
  Organic Photovoltaic Devices. \emph{Adv. Energy Mater.} \textbf{2021},
  \emph{11}, 2100342\relax
\mciteBstWouldAddEndPuncttrue
\mciteSetBstMidEndSepPunct{\mcitedefaultmidpunct}
{\mcitedefaultendpunct}{\mcitedefaultseppunct}\relax
\EndOfBibitem
\bibitem[Jena \latin{et~al.}(2019)Jena, Kulkarni, and Miyasaka]{perov_pv1}
Jena,~A.~K.; Kulkarni,~A.; Miyasaka,~T. Halide Perovskite Photovoltaics:
  Background, Status, and Future Prospects. \emph{Chem. Rev.} \textbf{2019},
  \emph{119}, 3036--3103\relax
\mciteBstWouldAddEndPuncttrue
\mciteSetBstMidEndSepPunct{\mcitedefaultmidpunct}
{\mcitedefaultendpunct}{\mcitedefaultseppunct}\relax
\EndOfBibitem
\bibitem[Wells(1893)]{first_perov}
Wells,~H.~L. Über die Cäsium- und Kalium-Bleihalogenide. \emph{Z. Anorg.
  Allg. Chem.} \textbf{1893}, \emph{3}, 195--210\relax
\mciteBstWouldAddEndPuncttrue
\mciteSetBstMidEndSepPunct{\mcitedefaultmidpunct}
{\mcitedefaultendpunct}{\mcitedefaultseppunct}\relax
\EndOfBibitem
\bibitem[Xu \latin{et~al.}(2023)Xu, Zhang, Li, Yang, and Zhu]{perov_pv2}
Xu,~F.; Zhang,~M.; Li,~Z.; Yang,~X.; Zhu,~R. Challenges and Perspectives toward
  Future Wide-Bandgap Mixed-Halide Perovskite Photovoltaics. \emph{Adv. Energy
  Mater.} \textbf{2023}, \emph{13}, 2203911\relax
\mciteBstWouldAddEndPuncttrue
\mciteSetBstMidEndSepPunct{\mcitedefaultmidpunct}
{\mcitedefaultendpunct}{\mcitedefaultseppunct}\relax
\EndOfBibitem
\bibitem[Wang \latin{et~al.}(2021)Wang, Huang, Xue, Tong, Zhu, and
  Yang]{perov_pv3}
Wang,~R.; Huang,~T.; Xue,~J.; Tong,~J.; Zhu,~K.; Yang,~Y. Prospects for Metal
  Halide Perovskite-Based Tandem Solar Cells. \emph{Nat. Photon.}
  \textbf{2021}, \emph{15}, 411--425\relax
\mciteBstWouldAddEndPuncttrue
\mciteSetBstMidEndSepPunct{\mcitedefaultmidpunct}
{\mcitedefaultendpunct}{\mcitedefaultseppunct}\relax
\EndOfBibitem
\bibitem[Chen \latin{et~al.}(2021)Chen, Liu, Wang, and Zhao]{perov_pv4}
Chen,~Y.; Liu,~X.; Wang,~T.; Zhao,~Y. Highly Stable Inorganic Lead Halide
  Perovskite toward Efficient Photovoltaics. \emph{Acc. Chem. Res.}
  \textbf{2021}, \emph{54}, 3452--3461, PMID: 34428021\relax
\mciteBstWouldAddEndPuncttrue
\mciteSetBstMidEndSepPunct{\mcitedefaultmidpunct}
{\mcitedefaultendpunct}{\mcitedefaultseppunct}\relax
\EndOfBibitem
\bibitem[Yang \latin{et~al.}(2020)Yang, Igbari, Lou, Wang, and Liao]{perov_pv5}
Yang,~W.-F.; Igbari,~F.; Lou,~Y.-H.; Wang,~Z.-K.; Liao,~L.-S. Tin Halide
  Perovskites: Progress and Challenges. \emph{Adv. Energy Mater.}
  \textbf{2020}, \emph{10}, 1902584\relax
\mciteBstWouldAddEndPuncttrue
\mciteSetBstMidEndSepPunct{\mcitedefaultmidpunct}
{\mcitedefaultendpunct}{\mcitedefaultseppunct}\relax
\EndOfBibitem
\bibitem[Tong \latin{et~al.}(2022)Tong, Najar, Wang, Liu, Du, Yang, Li, Wang,
  and Liu]{tunable_band1}
Tong,~Y.; Najar,~A.; Wang,~L.; Liu,~L.; Du,~M.; Yang,~J.; Li,~J.; Wang,~K.;
  Liu,~S.~F. Wide-Bandgap Organic–Inorganic Lead Halide Perovskite Solar
  Cells. \emph{Adv. Sci.} \textbf{2022}, \emph{9}, 2105085\relax
\mciteBstWouldAddEndPuncttrue
\mciteSetBstMidEndSepPunct{\mcitedefaultmidpunct}
{\mcitedefaultendpunct}{\mcitedefaultseppunct}\relax
\EndOfBibitem
\bibitem[Gu \latin{et~al.}(2020)Gu, Lin, Han, Gao, Tan, and Zhu]{tunable_band2}
Gu,~S.; Lin,~R.; Han,~Q.; Gao,~Y.; Tan,~H.; Zhu,~J. Tin and Mixed Lead–Tin
  Halide Perovskite Solar Cells: Progress and their Application in Tandem Solar
  Cells. \emph{Adv. Mater.} \textbf{2020}, \emph{32}, 1907392\relax
\mciteBstWouldAddEndPuncttrue
\mciteSetBstMidEndSepPunct{\mcitedefaultmidpunct}
{\mcitedefaultendpunct}{\mcitedefaultseppunct}\relax
\EndOfBibitem
\bibitem[Rigter \latin{et~al.}(2021)Rigter, Quinn, Kumar, Fenning, Massonnet,
  Ellis, Heeren, Svane, Walsh, and Garnett]{tunable_band3}
Rigter,~S.~A.; Quinn,~X.~L.; Kumar,~R.~E.; Fenning,~D.~P.; Massonnet,~P.;
  Ellis,~S.~R.; Heeren,~R. M.~A.; Svane,~K.~L.; Walsh,~A.; Garnett,~E.~C.
  Passivation Properties and Formation Mechanism of Amorphous Halide Perovskite
  Thin Films. \emph{Adv. Funct. Mater.} \textbf{2021}, \emph{31}, 2010330\relax
\mciteBstWouldAddEndPuncttrue
\mciteSetBstMidEndSepPunct{\mcitedefaultmidpunct}
{\mcitedefaultendpunct}{\mcitedefaultseppunct}\relax
\EndOfBibitem
\bibitem[He \latin{et~al.}(2022)He, Wu, Zhang, Wang, Fu, Liu, and
  Zhong]{PhysRevB.105.064104}
He,~R.; Wu,~H.; Zhang,~L.; Wang,~X.; Fu,~F.; Liu,~S.; Zhong,~Z. Structural
  phase transitions in $\mathrm{SrTi}{\mathrm{O}}_{3}$ from deep potential
  molecular dynamics. \emph{Phys. Rev. B} \textbf{2022}, \emph{105},
  064104\relax
\mciteBstWouldAddEndPuncttrue
\mciteSetBstMidEndSepPunct{\mcitedefaultmidpunct}
{\mcitedefaultendpunct}{\mcitedefaultseppunct}\relax
\EndOfBibitem
\bibitem[Jain \latin{et~al.}(2019)Jain, Chaube, Khullar, Goverapet~Srinivasan,
  and Rai]{C9CP03240A}
Jain,~D.; Chaube,~S.; Khullar,~P.; Goverapet~Srinivasan,~S.; Rai,~B. Bulk and
  surface DFT investigations of inorganic halide perovskites screened using
  machine learning and materials property databases. \emph{Phys. Chem. Chem.
  Phys.} \textbf{2019}, \emph{21}, 19423--19436\relax
\mciteBstWouldAddEndPuncttrue
\mciteSetBstMidEndSepPunct{\mcitedefaultmidpunct}
{\mcitedefaultendpunct}{\mcitedefaultseppunct}\relax
\EndOfBibitem
\bibitem[Grote and Berger(2015)Grote, and Berger]{pbe1}
Grote,~C.; Berger,~R.~F. Strain Tuning of Tin–Halide and Lead–Halide
  Perovskites: A First-Principles Atomic and Electronic Structure Study.
  \emph{J. Phys. Chem. C} \textbf{2015}, \emph{119}, 22832--22837\relax
\mciteBstWouldAddEndPuncttrue
\mciteSetBstMidEndSepPunct{\mcitedefaultmidpunct}
{\mcitedefaultendpunct}{\mcitedefaultseppunct}\relax
\EndOfBibitem
\bibitem[ten Brinck and Infante(2016)ten Brinck, and Infante]{pbe2}
ten Brinck,~S.; Infante,~I. Surface Termination, Morphology, and Bright
  Photoluminescence of Cesium Lead Halide Perovskite Nanocrystals. \emph{ACS
  Energy Lett.} \textbf{2016}, \emph{1}, 1266--1272\relax
\mciteBstWouldAddEndPuncttrue
\mciteSetBstMidEndSepPunct{\mcitedefaultmidpunct}
{\mcitedefaultendpunct}{\mcitedefaultseppunct}\relax
\EndOfBibitem
\bibitem[Kashtiban \latin{et~al.}(2023)Kashtiban, Patrick, Ramasse, Walton, and
  Sloan]{pbe3}
Kashtiban,~R.~J.; Patrick,~C.~E.; Ramasse,~Q.; Walton,~R.~I.; Sloan,~J.
  Picoperovskites: The Smallest Conceivable Isolated Halide Perovskite
  Structures Formed within Carbon Nanotubes. \emph{Adv. Mater.} \textbf{2023},
  \emph{35}, 2208575\relax
\mciteBstWouldAddEndPuncttrue
\mciteSetBstMidEndSepPunct{\mcitedefaultmidpunct}
{\mcitedefaultendpunct}{\mcitedefaultseppunct}\relax
\EndOfBibitem
\bibitem[Xiao and Yan(2017)Xiao, and
  Yan]{https://doi.org/10.1002/aenm.201701136}
Xiao,~Z.; Yan,~Y. Progress in Theoretical Study of Metal Halide Perovskite
  Solar Cell Materials. \emph{Adv. Energy Mater.} \textbf{2017}, \emph{7},
  1701136\relax
\mciteBstWouldAddEndPuncttrue
\mciteSetBstMidEndSepPunct{\mcitedefaultmidpunct}
{\mcitedefaultendpunct}{\mcitedefaultseppunct}\relax
\EndOfBibitem
\bibitem[Brivio \latin{et~al.}(2014)Brivio, Butler, Walsh, and van
  Schilfgaarde]{PhysRevB.89.155204}
Brivio,~F.; Butler,~K.~T.; Walsh,~A.; van Schilfgaarde,~M. Relativistic
  quasiparticle self-consistent electronic structure of hybrid halide
  perovskite photovoltaic absorbers. \emph{Phys. Rev. B} \textbf{2014},
  \emph{89}, 155204\relax
\mciteBstWouldAddEndPuncttrue
\mciteSetBstMidEndSepPunct{\mcitedefaultmidpunct}
{\mcitedefaultendpunct}{\mcitedefaultseppunct}\relax
\EndOfBibitem
\bibitem[Even \latin{et~al.}(2013)Even, Pedesseau, Jancu, and
  Katan]{doi:10.1021/jz401532q}
Even,~J.; Pedesseau,~L.; Jancu,~J.-M.; Katan,~C. Importance of Spin–Orbit
  Coupling in Hybrid Organic/Inorganic Perovskites for Photovoltaic
  Applications. \emph{J. Phys. Chem. Lett.} \textbf{2013}, \emph{4},
  2999--3005\relax
\mciteBstWouldAddEndPuncttrue
\mciteSetBstMidEndSepPunct{\mcitedefaultmidpunct}
{\mcitedefaultendpunct}{\mcitedefaultseppunct}\relax
\EndOfBibitem
\bibitem[Wiktor \latin{et~al.}(2017)Wiktor, Rothlisberger, and
  Pasquarello]{doi:10.1021/acs.jpclett.7b02648}
Wiktor,~J.; Rothlisberger,~U.; Pasquarello,~A. Predictive Determination of Band
  Gaps of Inorganic Halide Perovskites. \emph{J. Phys. Chem. Lett.}
  \textbf{2017}, \emph{8}, 5507--5512, PMID: 29077408\relax
\mciteBstWouldAddEndPuncttrue
\mciteSetBstMidEndSepPunct{\mcitedefaultmidpunct}
{\mcitedefaultendpunct}{\mcitedefaultseppunct}\relax
\EndOfBibitem
\bibitem[Kaiser \latin{et~al.}(2021)Kaiser, Carignano, Alothman, Mosconi,
  Kachmar, Goddard, and De~Angelis]{doi:10.1021/acs.jpclett.1c03428}
Kaiser,~W.; Carignano,~M.; Alothman,~A.~A.; Mosconi,~E.; Kachmar,~A.;
  Goddard,~W. A.~I.; De~Angelis,~F. First-Principles Molecular Dynamics in
  Metal-Halide Perovskites: Contrasting Generalized Gradient Approximation and
  Hybrid Functionals. \emph{J. Phys. Chem. Lett.} \textbf{2021}, \emph{12},
  11886--11893, PMID: 34875174\relax
\mciteBstWouldAddEndPuncttrue
\mciteSetBstMidEndSepPunct{\mcitedefaultmidpunct}
{\mcitedefaultendpunct}{\mcitedefaultseppunct}\relax
\EndOfBibitem
\bibitem[Chen \latin{et~al.}(2020)Chen, Zhang, Wang, and E]{chen2020ground}
Chen,~Y.; Zhang,~L.; Wang,~H.; E,~W. Ground State Energy Functional with
  Hartree--Fock Efficiency and Chemical Accuracy. \emph{J. Phys. Chem. A}
  \textbf{2020}, \emph{124}, 7155--7165\relax
\mciteBstWouldAddEndPuncttrue
\mciteSetBstMidEndSepPunct{\mcitedefaultmidpunct}
{\mcitedefaultendpunct}{\mcitedefaultseppunct}\relax
\EndOfBibitem
\bibitem[Chen \latin{et~al.}(2020)Chen, Zhang, Wang, and E]{chen2020deepks}
Chen,~Y.; Zhang,~L.; Wang,~H.; E,~W. DeePKS: a Comprehensive Data-Driven
  Approach towards Chemically Accurate Density Functional Theory. \emph{J.
  Chem. Theory Comput.} \textbf{2020}, \emph{17}, 170--181\relax
\mciteBstWouldAddEndPuncttrue
\mciteSetBstMidEndSepPunct{\mcitedefaultmidpunct}
{\mcitedefaultendpunct}{\mcitedefaultseppunct}\relax
\EndOfBibitem
\bibitem[Li \latin{et~al.}(2022)Li, Ou, Chen, Cao, Liu, Zhang, Zheng, Cai, Wu,
  Wang, Chen, and Zhang]{deepks_abacus}
Li,~W.; Ou,~Q.; Chen,~Y.; Cao,~Y.; Liu,~R.; Zhang,~C.; Zheng,~D.; Cai,~C.;
  Wu,~X.; Wang,~H.; Chen,~M.; Zhang,~L. DeePKS + ABACUS as a Bridge between
  Expensive Quantum Mechanical Models and Machine Learning Potentials. \emph{J.
  Phys. Chem. A} \textbf{2022}, \emph{126}, 9154--9164, PMID: 36455227\relax
\mciteBstWouldAddEndPuncttrue
\mciteSetBstMidEndSepPunct{\mcitedefaultmidpunct}
{\mcitedefaultendpunct}{\mcitedefaultseppunct}\relax
\EndOfBibitem
\bibitem[Das \latin{et~al.}(2022)Das, Di~Liberto, and
  Pacchioni]{doi:10.1021/acs.jpcc.1c09594}
Das,~T.; Di~Liberto,~G.; Pacchioni,~G. Density Functional Theory Estimate of
  Halide Perovskite Band Gap: When Spin Orbit Coupling Helps. \emph{J. Phys.
  Chem. C} \textbf{2022}, \emph{126}, 2184--2198\relax
\mciteBstWouldAddEndPuncttrue
\mciteSetBstMidEndSepPunct{\mcitedefaultmidpunct}
{\mcitedefaultendpunct}{\mcitedefaultseppunct}\relax
\EndOfBibitem
\bibitem[Li \latin{et~al.}(2016)Li, Liu, Chen, Lin, Ren, Lin, Yang, and
  He]{li_abacus}
Li,~P.; Liu,~X.; Chen,~M.; Lin,~P.; Ren,~X.; Lin,~L.; Yang,~C.; He,~L.
  Large-Scale $Ab\ Initio$ Simulations Based on Systematically Improvable
  Atomic Basis. \emph{Comput. Mat. Sci.} \textbf{2016}, \emph{112},
  503--517\relax
\mciteBstWouldAddEndPuncttrue
\mciteSetBstMidEndSepPunct{\mcitedefaultmidpunct}
{\mcitedefaultendpunct}{\mcitedefaultseppunct}\relax
\EndOfBibitem
\bibitem[Chen \latin{et~al.}(2010)Chen, Guo, and He]{chen_abacus}
Chen,~M.; Guo,~G.-C.; He,~L. Systematically Improvable Optimized Atomic Basis
  Sets for $Ab\ Initio$ Calculations. \emph{J. Phys.: Condens. Matt.}
  \textbf{2010}, \emph{22}, 445501\relax
\mciteBstWouldAddEndPuncttrue
\mciteSetBstMidEndSepPunct{\mcitedefaultmidpunct}
{\mcitedefaultendpunct}{\mcitedefaultseppunct}\relax
\EndOfBibitem
\bibitem[Tuo \latin{et~al.}()Tuo, Li, Wang, Chen, Zhong, Xu, and
  Dai]{https://doi.org/10.1002/adfm.202301663}
Tuo,~P.; Li,~L.; Wang,~X.; Chen,~J.; Zhong,~Z.; Xu,~B.; Dai,~F.-Z. Spontaneous
  Hybrid Nano-Domain Behavior of the Organic–Inorganic Hybrid Perovskites.
  \emph{Adv. Funct. Mater.} \emph{n/a}, 2301663\relax
\mciteBstWouldAddEndPuncttrue
\mciteSetBstMidEndSepPunct{\mcitedefaultmidpunct}
{\mcitedefaultendpunct}{\mcitedefaultseppunct}\relax
\EndOfBibitem
\bibitem[Kresse and Furthmüller(1996)Kresse, and Furthmüller]{vasp1}
Kresse,~G.; Furthmüller,~J. Efficiency of ab-initio total energy calculations
  for metals and semiconductors using a plane-wave basis set. \emph{Comput.
  Mater. Sci.} \textbf{1996}, \emph{6}, 15--50\relax
\mciteBstWouldAddEndPuncttrue
\mciteSetBstMidEndSepPunct{\mcitedefaultmidpunct}
{\mcitedefaultendpunct}{\mcitedefaultseppunct}\relax
\EndOfBibitem
\bibitem[Kresse and Furthm\"uller(1996)Kresse, and Furthm\"uller]{vasp2}
Kresse,~G.; Furthm\"uller,~J. Efficient iterative schemes for ab initio
  total-energy calculations using a plane-wave basis set. \emph{Phys. Rev. B}
  \textbf{1996}, \emph{54}, 11169--11186\relax
\mciteBstWouldAddEndPuncttrue
\mciteSetBstMidEndSepPunct{\mcitedefaultmidpunct}
{\mcitedefaultendpunct}{\mcitedefaultseppunct}\relax
\EndOfBibitem
\bibitem[Kresse and Hafner(1993)Kresse, and Hafner]{vasp3}
Kresse,~G.; Hafner,~J. Ab initio molecular dynamics for open-shell transition
  metals. \emph{Phys. Rev. B} \textbf{1993}, \emph{48}, 13115--13118\relax
\mciteBstWouldAddEndPuncttrue
\mciteSetBstMidEndSepPunct{\mcitedefaultmidpunct}
{\mcitedefaultendpunct}{\mcitedefaultseppunct}\relax
\EndOfBibitem
\bibitem[Yang \latin{et~al.}(2022)Yang, Li, Chen, Feng, Wu, Gates, Gao, Ding,
  Yao, and Li]{cspbi3}
Yang,~W.; Li,~J.; Chen,~X.; Feng,~Y.; Wu,~C.; Gates,~I.~D.; Gao,~Z.; Ding,~X.;
  Yao,~J.; Li,~H. Exploring the Effects of Ionic Defects on the Stability of
  CsPbI3 with a Deep Learning Potential. \emph{Chem. Phys. Chem.}
  \textbf{2022}, \emph{23}, e202100841\relax
\mciteBstWouldAddEndPuncttrue
\mciteSetBstMidEndSepPunct{\mcitedefaultmidpunct}
{\mcitedefaultendpunct}{\mcitedefaultseppunct}\relax
\EndOfBibitem
\bibitem[Jin \latin{et~al.}(2021)Jin, Zheng, and He]{lcao1}
Jin,~G.; Zheng,~D.; He,~L. Calculation of Berry Curvature Using Non-Orthogonal
  Atomic Orbitals. \emph{J. Phys. Condens. Matter.} \textbf{2021}, \emph{33},
  325503\relax
\mciteBstWouldAddEndPuncttrue
\mciteSetBstMidEndSepPunct{\mcitedefaultmidpunct}
{\mcitedefaultendpunct}{\mcitedefaultseppunct}\relax
\EndOfBibitem
\bibitem[Lin \latin{et~al.}(2021)Lin, Ren, and He]{lcao2}
Lin,~P.; Ren,~X.; He,~L. Strategy for Constructing Compact Numerical Atomic
  Orbital Basis Sets by Incorporating the Gradients of Reference Wavefunctions.
  \emph{Phys. Rev. B} \textbf{2021}, \emph{103}, 235131\relax
\mciteBstWouldAddEndPuncttrue
\mciteSetBstMidEndSepPunct{\mcitedefaultmidpunct}
{\mcitedefaultendpunct}{\mcitedefaultseppunct}\relax
\EndOfBibitem
\bibitem[Hamann(2013)]{oncv}
Hamann,~D.~R. Optimized Norm-Conserving Vanderbilt Pseudopotentials.
  \emph{Phys. Rev. B} \textbf{2013}, \emph{88}, 085117\relax
\mciteBstWouldAddEndPuncttrue
\mciteSetBstMidEndSepPunct{\mcitedefaultmidpunct}
{\mcitedefaultendpunct}{\mcitedefaultseppunct}\relax
\EndOfBibitem
\bibitem[Zhang \latin{et~al.}(2020)Zhang, Zhang, and
  Lu]{doi:10.1021/acs.jpclett.0c02135}
Zhang,~L.; Zhang,~X.; Lu,~G. Intramolecular Band Alignment and Spin–Orbit
  Coupling in Two-Dimensional Halide Perovskites. \emph{J. Phys. Chem. Lett.}
  \textbf{2020}, \emph{11}, 6982--6989, PMID: 32787199\relax
\mciteBstWouldAddEndPuncttrue
\mciteSetBstMidEndSepPunct{\mcitedefaultmidpunct}
{\mcitedefaultendpunct}{\mcitedefaultseppunct}\relax
\EndOfBibitem
\bibitem[Yumoto \latin{et~al.}(2021)Yumoto, Hirori, Sekiguchi, Sato, Saruyama,
  Teranishi, and Kanemitsu]{lead_soc_nc}
Yumoto,~G.; Hirori,~H.; Sekiguchi,~F.; Sato,~R.; Saruyama,~M.; Teranishi,~T.;
  Kanemitsu,~Y. Strong Spin-Orbit Coupling Inducing Autler-Townes Effect in
  Lead Halide Perovskite Nanocrystals. \emph{Nat. Commun.} \textbf{2021},
  \emph{12}\relax
\mciteBstWouldAddEndPuncttrue
\mciteSetBstMidEndSepPunct{\mcitedefaultmidpunct}
{\mcitedefaultendpunct}{\mcitedefaultseppunct}\relax
\EndOfBibitem
\bibitem[Chatterjee \latin{et~al.}(2020)Chatterjee, Payne, Irvine, and
  Pal]{C9TA12263J}
Chatterjee,~S.; Payne,~J.; Irvine,~J. T.~S.; Pal,~A.~J. Bandgap bowing in a
  zero-dimensional hybrid halide perovskite derivative: spin–orbit coupling
  versus lattice strain. \emph{J. Mater. Chem. A} \textbf{2020}, \emph{8},
  4416--4427\relax
\mciteBstWouldAddEndPuncttrue
\mciteSetBstMidEndSepPunct{\mcitedefaultmidpunct}
{\mcitedefaultendpunct}{\mcitedefaultseppunct}\relax
\EndOfBibitem
\bibitem[Scherpelz \latin{et~al.}(2016)Scherpelz, Govoni, Hamada, and
  Galli]{doi:10.1021/acs.jctc.6b00114}
Scherpelz,~P.; Govoni,~M.; Hamada,~I.; Galli,~G. Implementation and Validation
  of Fully Relativistic GW Calculations: Spin–Orbit Coupling in Molecules,
  Nanocrystals, and Solids. \emph{J. Chem. Theory Comput.} \textbf{2016},
  \emph{12}, 3523--3544, PMID: 27331614\relax
\mciteBstWouldAddEndPuncttrue
\mciteSetBstMidEndSepPunct{\mcitedefaultmidpunct}
{\mcitedefaultendpunct}{\mcitedefaultseppunct}\relax
\EndOfBibitem
\bibitem[Hafner(2008)]{https://doi.org/10.1002/jcc.21057}
Hafner,~J. Ab-initio simulations of materials using VASP: Density-functional
  theory and beyond. \emph{J. Comput. Chem.} \textbf{2008}, \emph{29},
  2044--2078\relax
\mciteBstWouldAddEndPuncttrue
\mciteSetBstMidEndSepPunct{\mcitedefaultmidpunct}
{\mcitedefaultendpunct}{\mcitedefaultseppunct}\relax
\EndOfBibitem
\end{mcitethebibliography}

\end{document}


\begin{table}[t]
\centering
\caption{The contribution of each element to the absolute energy difference between the base and target method for all involved halide perovskites. Numbers are shown in the unit of a.u. Note that such large absolute energy difference is result from the fact that the SCF calculation of target functional HSE06 is performed with the PAW method while that of the base method PBE is performed with the SG15 ONCV pseudopotentials.}
\begin{tabular}{cc}
\toprule
element $I$ &  $\epsilon_I$ \\ 
\midrule
H	& 4.3297 \\
C	& 2.5416  \\
N	& 13.4613  \\
Cl	& 18.9693  \\
Br	& 17.4021  \\
Sn	& 27.3945   \\
I	& 105.2054 \\
Cs	& 44.6507 \\
Pb	& 19.7978  \\
\bottomrule
\end{tabular}
\end{table}

\begin{landscape}
\begin{table}[htp]
\centering
\caption{Training sets mean absolute errors (MAE) of energy, force, stress, and band gap given by DeePKS and PBE with respect to the HSE06 functional. DeePKS and PBE calculations are performed in ABACUS with numerical atomic orbital (NAO) basis set and SG15 ONCV pseudopotentials while HSE06 calculations are performed in VASP with plane-wave basis and PAW method.}
\begin{tabular}{ccccccccccc}
\toprule
\multicolumn{3}{c}{MAE:} & \multicolumn{2}{c}{Energy (meV/atom)} & \multicolumn{2}{c}{Force (eV/$\mathring{\textrm{A}}$)}  & \multicolumn{2}{c}{Stress (eV)} & \multicolumn{2}{c}{Band gap (eV)} \\
\midrule
systems	& \# of atoms	&	\# of frames	& DeePKS	&	PBE	&	DeePKS	&	PBE	&	DeePKS	&	PBE	&	DeePKS	&	PBE	\\
\midrule
FAPbI$_3$	&	48	&	48	&	0.6805	&	1.6230	&	0.0551	&	0.1104	&	0.2297	&	3.1150	&	0.0355	&	0.7207	\\
FAPbI$_3$	&	24	&	41	&	1.1904	&	2.2587	&	0.0651	&	0.1197	&	0.2677	&	1.6096	&	0.0672	&	0.8276	\\
MAPbI$_3$	&	48	&	78	&	0.7200	&	1.3188	&	0.0460	&	0.0998	&	0.2809	&	2.2601	&	0.0339	&	0.7144	\\
MAPbI$_3$	&	96	&	13	&	0.9851	&	1.0110	&	0.0557	&	0.1056	&	0.7897	&	4.5385	&	0.1049	&	0.7629	\\
CsPbI$_3$	&	10	&	6	&	0.1590	&	1.1691	&	0.0062	&	0.0036	&	0.0291	&	0.4093	&	0.0369	&	0.6716	\\
CsPbI$_3$	&	20	&	61	&	2.1979	&	3.3897	&	0.0257	&	0.0498	&	0.1536	&	0.8712	&	0.1009	&	0.7859	\\
CsPbI$_3$	&	40	&	8	&	1.1791	&	2.2761	&	0.0129	&	0.0224	&	0.1569	&	1.6445	&	0.0387	&	0.6062	\\
CsPbCl$_3$	&	5	&	45	&	2.1649	&	4.7158	&	0.0175	&	0.0461	&	0.0688	&	0.2760	&	0.0505	&	1.0011	\\
CsSnBr$_3$	&	5	&	44	&	2.1258	&	11.6558	&	0.0162	&	0.0644	&	0.0477	&	0.3085	&	0.0394	&	0.8448	\\
MAPbBr$_3$	&	12	&	77	&	1.8080	&	3.1812	&	0.0434	&	0.1011	&	0.0953	&	0.6994	&	0.0554	&	0.9077	\\
MASnCl$_3$	&	12	&	39	&	1.6948	&	3.6568	&	0.0453	&	0.1147	&	0.0997	&	0.7624	&	0.0645	&	0.8668	\\
\midrule
\multicolumn{3}{c}{Overall MAE:} & 1.5025	&	3.5460	&	0.0390	&	0.0864	&	0.1770	&	1.3677	&	0.0564	&	0.8193\\
\bottomrule
\end{tabular}
\end{table}

\begin{table}[htp]
\centering
\caption{Test sets mean absolute errors (MAE) of energy, force, stress, and band gap given by DeePKS and PBE with respect to the HSE06 functional. DeePKS and PBE calculations are performed in ABACUS with numerical atomic orbital (NAO) basis set and SG15 ONCV pseudopotentials while HSE06 calculations are performed in VASP with plane-wave basis and PAW method.}
\begin{tabular}{ccccccccccc}
\toprule
\multicolumn{3}{c}{MAE:} & \multicolumn{2}{c}{Energy (meV/atom)} & \multicolumn{2}{c}{Force (eV/$\mathring{\textrm{A}}$)}  & \multicolumn{2}{c}{Stress (eV)} & \multicolumn{2}{c}{Band gap (eV)} \\
\midrule
systems	& \# of atoms	&	\# of frames	& DeePKS	&	PBE	&	DeePKS	&	PBE	&	DeePKS	&	PBE	&	DeePKS	&	PBE	\\
\midrule
CsSnI$_3$	&	5	&	81	&	4.0443	&	5.7440	&	0.0357	&	0.0609	&	0.1287	&	0.2442	&	0.0504	&	0.6690	\\
MAPbCl$_3$	&	12	&	131	&	1.8487	&	2.7568	&	0.0505	&	0.0994	&	0.1201	&	0.6913	&	0.0620	&	0.9848	\\
MASnBr$_3$	&	12	&	150	&	1.6353	&	3.8258	&	0.0452	&	0.1083	&	0.1145	&	0.7269	&	0.0488	&	0.7800	\\
FA$_{0.125}$Cs$_{0.875}$PbI$_3$	&	47	&	11	&	0.0320	&	1.4994	&	0.0336	&	0.0522	&	0.2463	&	2.3415	&	0.0223	&	0.6965	\\
\midrule
\multicolumn{3}{c}{Overall MAE:} & 2.1861	&	3.7983	&	0.0447	&	0.0932	&	0.1234	&	0.6572	&	0.0530	&	0.8254\\
\bottomrule
\end{tabular}
\end{table}

\end{landscape}

\begin{table}
\centering
\caption{Total CPU times (in the unit of second) of the SCF calculation performed via HSE06, DeePKS, and PBE for all tested perovskites. All tested non-hybrid perovskites are cubic phase except for those indicated by Greek letters. The HSE06 calculations are carried out in VASP version 5.4 with plane-wave basis and PAW method on 16-core 256-mem CPU processor, starting from the converged PBE wavefunction. DeePKS and PBE calculations are carried out in ABACUS version 3.2 with DZP NAO basis set and SG15 ONCV pseudopotentials on 16-core 64-mem CPU processor. All calculations are ran with 16-core MPI parallelization. Input parameters of HSE06 (in VASP), DeePKS (in ABACUS), and PBE (in ABACUS) are provided below.}
\begin{tabular}{cccc}
\toprule
system & HSE06 & DeePKS & PBE \\
\midrule
CsPbBr$_3$	&	111.39012	&	69.45452	&	16875.736	\\
CsPbCl$_3$	&	143.67532	&	58.22072	&	14457.284	\\
CsPbI$_3$	&	195.44233	&	94.71545	&	13227.812	\\
$\beta$-CsPbI$_3$	&	255.41758	&	107.91931	&	192627.938	\\
$\gamma$-CsPbI$_3$	&	331.70495	&	182.74886	&	250612.344	\\
$\delta$-CsPbI$_3$	&	1473.79506	&	237.27452	&	300858.344	\\
CsSnBr$_3$	&	97.78578	&	57.63274	&	15659.790	\\
CsSnCl$_3$	&	155.86923	&	58.71236	&	14514.202	\\
CsSnI$_3$	&	193.96808	&	92.26651	&	19305.125	\\
FAPbBr$_3$	&	233.16509	&	134.81837	&	176083.594	\\
FAPbCl$_3$	&	334.87374	&	134.87047	&	155415.516	\\
FAPbI$_3$	&	310.39089	&	174.69117	&	345250.188	\\
FASnBr$_3$	&	217.74080	&	111.85347	&	163583.922	\\
FASnCl$_3$	&	269.62527	&	128.92052	&	172128.469	\\
FASnI$_3$	&	316.82600	&	156.07270	&	189681.109	\\
MAPbBr$_3$	&	254.16727	&	119.34650	&	136092.531	\\
MAPbCl$_3$	&	323.73752	&	125.74404	&	97563.695	\\
MAPbI$_3$	&	379.85375	&	154.58523	&	220930.016	\\
MASnBr$_3$	&	288.36057	&	105.26187	&	115660.477	\\
MASnCl$_3$	&	282.27638	&	114.72428	&	114202.844	\\
MASnI$_3$	&	451.40572	&	134.93733	&	113742.766	\\
FA$_{0.125}$Cs$_{0.875}$PbI$_3$	&	1072.25264	&	375.18319	&	774865.938	\\
FA$_{0.25}$Cs$_{0.75}$PbI$_3$	&	873.54869	&	463.76868	&	632852.938	\\
FA$_{0.375}$Cs$_{0.625}$PbI$_3$	&	736.42967	&	653.44151	&	689350.938	\\
FA$_{0.5}$Cs$_{0.5}$PbI$_3$	&	1107.17502	&	616.79290	&	822730.000	\\
FA$_{0.625}$Cs$_{0.375}$PbI$_3$	&	2049.49370	&	555.79434	&	805206.375	\\
FA$_{0.75}$Cs$_{0.25}$PbI$_3$	&	1251.96995	&	587.64888	&	791275.375	\\
FA$_{0.875}$Cs$_{0.125}$PbI$_3$	&	1124.27960	&	611.86466	&	775134.500	\\
\bottomrule
\end{tabular}
\end{table}

\newpage
\section{Input parameters of HSE06, DeePKS, and PBE SCF calculations.}
\subsection{HSE06 jobs in VASP}
\begin{spverbatim}
 SYSTEM = perovskite
 ISTART = 1 //starts from a converged PBE wavefunction
 ICHARG=0
 ENCUT  =    500 // in the unit of eV
 PREC = Accurate
 ISMEAR = 0 // Gaussian smearing
 SIGMA = 0.02 // in the unit of eV
 EDIFF = 1E-05 
 GGA = PE
 LREAL= Auto
 LORBIT = 11
 ISIF = 2

 LHFCALC = .TRUE.
 ALGO = Damped
 TIME = 0.4
 HFSCREEN = 0.2
 PRECFOCK = Fast
 AEXX = 0.25
 NELM = 200
\end{spverbatim}

\subsection{DeePKS/PBE jobs in ABACUS}
\begin{spverbatim}
 INPUT_PARAMETERS
 calculation scf
 ecutwfc 100.000000   // energy cutoff in the unit of Rydberg
 scf_thr 1.000000e-07 
 scf_nmax 100
 basis_type lcao
 dft_functional pbe
 gamma_only 0
 mixing_type pulay
 mixing_beta 0.400000
 out_dos 1
 nspin 1
 smearing_method gaussian 
 smearing_sigma 0.00147  //in the unit of Rydberg
 cal_force 1
 cal_stress 1
 out_bandgap 1
 deepks_scf 1     // 1 for DeePKS; 0 for PBE
 deepks_model ./model.ptg
\end{spverbatim}

\begin{longtable}{ccccccc}
    \caption{Stress elements predicted by HSE06, DeePKS, and PBE for hybrid perovskites. Absolute errors of the DeePKS and PBE results with respect to HSE06 results are listed. Numbers are shown in the unit of KBar.}    \\
    \toprule
system	&	stress	&	HSE06	&	DeePKS	&	DeePKS &	PBE	&	PBE \\
 & element & & & Abs. Err. & & Abs. Err. \\
 \midrule
	&	$\sigma_{xx}$	&	1.9699	&	2.2348	&	0.2649	&	5.4021	&	3.4322	\\
	&	$\sigma_{yy}$	&	2.0891	&	2.8939	&	0.8049	&	4.7289	&	2.6398	\\
FA$_{0.125}$Cs$_{0.875}$PbI$_3$	&	$\sigma_{zz}$	&	1.7496	&	2.4178	&	0.6682	&	5.0406	&	3.2909	\\
	&	$\sigma_{xy}$	&	-0.5075	&	-0.6343	&	0.1268	&	0.0848	&	0.5923	\\
	&	$\sigma_{xz}$	&	-0.1183	&	-0.2782	&	0.1598	&	-0.0087	&	0.1096	\\
	&	$\sigma_{yz}$	&	0.2939	&	0.3603	&	0.0664	&	-0.0731	&	0.3670	\\
 \midrule
	&	$\sigma_{xx}$	&	0.4948	&	0.8636	&	0.3688	&	5.9766	&	5.4818	\\
	&	$\sigma_{yy}$	&	2.8390	&	3.3634	&	0.5244	&	6.5914	&	3.7525	\\
FA$_{0.25}$Cs$_{0.75}$PbI$_3$	&	$\sigma_{zz}$	&	1.8912	&	2.3546	&	0.4635	&	5.6441	&	3.7529	\\
	&	$\sigma_{xy}$	&	0.3537	&	0.3457	&	0.0080	&	0.1304	&	0.2233	\\
	&	$\sigma_{xz}$	&	0.1225	&	0.0297	&	0.0928	&	-0.0524	&	0.1750	\\
	&	$\sigma_{yz}$	&	0.0857	&	0.1056	&	0.0199	&	-0.1187	&	0.2044	\\
 \midrule
	&	$\sigma_{xx}$	&	1.9872	&	2.2120	&	0.2248	&	6.3285	&	4.3413	\\
	&	$\sigma_{yy}$	&	2.6901	&	3.1381	&	0.4481	&	6.8935	&	4.2034	\\
FA$_{0.375}$Cs$_{0.625}$PbI$_3$	&	$\sigma_{zz}$	&	2.4684	&	2.9213	&	0.4529	&	6.7989	&	4.3305	\\
	&	$\sigma_{xy}$	&	1.4172	&	1.2193	&	0.1979	&	-0.1119	&	1.5292	\\
	&	$\sigma_{xz}$	&	1.2061	&	1.0910	&	0.1151	&	-0.1455	&	1.3517	\\
	&	$\sigma_{yz}$	&	1.5097	&	1.6905	&	0.1809	&	0.5757	&	0.9340	\\
 \midrule
 \endfirsthead
    \caption{Stress elements predicted by HSE06, DeePKS, and PBE for hybrid perovskites. Absolute errors of the DeePKS and PBE results with respect to HSE06 results are listed. Numbers are shown in the unit of KBar. (cont.).}    \\
    \toprule
system	&	stress	&	HSE06	&	DeePKS	&	DeePKS &	PBE	&	PBE \\
 & element & & & Abs. Err. & & Abs. Err. \\
 \midrule
\endhead
	&	$\sigma_{xx}$	&	0.9048	&	1.1091	&	0.2043	&	7.3549	&	6.4501	\\
	&	$\sigma_{yy}$	&	0.3762	&	0.9593	&	0.5831	&	6.9435	&	6.5673	\\
FA$_{0.5}$Cs$_{0.5}$PbI$_3$	&	$\sigma_{zz}$	&	2.4929	&	2.9088	&	0.4159	&	7.4192	&	4.9262	\\
	&	$\sigma_{xy}$	&	1.2402	&	1.1141	&	0.1261	&	0.1268	&	1.1135	\\
	&	$\sigma_{xz}$	&	2.1786	&	1.9182	&	0.2604	&	0.1199	&	2.0587	\\
	&	$\sigma_{yz}$	&	0.7353	&	0.8147	&	0.0793	&	0.0340	&	0.7013	\\
 \midrule
	&	$\sigma_{xx}$	&	1.0742	&	1.3453	&	0.2711	&	7.1270	&	6.0528	\\
	&	$\sigma_{yy}$	&	1.5954	&	2.0233	&	0.4278	&	7.4609	&	5.8655	\\
FA$_{0.625}$Cs$_{0.375}$PbI$_3$	&	$\sigma_{zz}$	&	1.3722	&	1.6950	&	0.3228	&	7.3446	&	5.9724	\\
	&	$\sigma_{xy}$	&	2.5797	&	2.3406	&	0.2390	&	0.1039	&	2.4758	\\
	&	$\sigma_{xz}$	&	2.4207	&	2.2681	&	0.1527	&	0.0241	&	2.3966	\\
	&	$\sigma_{yz}$	&	2.0850	&	2.3039	&	0.2189	&	0.5464	&	1.5386	\\
 \midrule
	&	$\sigma_{xx}$	&	0.6725	&	0.9521	&	0.2796	&	7.5290	&	6.8565	\\
	&	$\sigma_{yy}$	&	1.0390	&	1.3473	&	0.3083	&	7.6636	&	6.6246	\\
FA$_{0.75}$Cs$_{0.25}$PbI$_3$	&	$\sigma_{zz}$	&	0.6870	&	0.8834	&	0.1964	&	7.5815	&	6.8944	\\
	&	$\sigma_{xy}$	&	3.1880	&	2.9220	&	0.2660	&	0.1630	&	3.0251	\\
	&	$\sigma_{xz}$	&	3.0759	&	2.9392	&	0.1367	&	0.1475	&	2.9284	\\
	&	$\sigma_{yz}$	&	2.2851	&	2.6175	&	0.3324	&	0.4901	&	1.7950	\\
 \midrule
	&	$\sigma_{xx}$	&	-0.0064	&	0.1278	&	0.1342	&	7.7951	&	7.8014	\\
	&	$\sigma_{yy}$	&	0.6432	&	1.1286	&	0.4853	&	7.9753	&	7.3321	\\
FA$_{0.875}$Cs$_{0.125}$PbI$_3$	&	$\sigma_{zz}$	&	0.3253	&	0.7446	&	0.4193	&	7.8975	&	7.5722	\\
	&	$\sigma_{xy}$	&	3.9724	&	3.6295	&	0.3429	&	0.3816	&	3.5907	\\
	&	$\sigma_{xz}$	&	3.7766	&	3.5036	&	0.2729	&	0.3293	&	3.4473	\\
	&	$\sigma_{yz}$	&	2.6741	&	2.8036	&	0.1295	&	0.5196	&	2.1545	\\
    \bottomrule
\endlastfoot
\end{longtable}

\clearpage
\begin{figure}[h]
  \centering
  \includegraphics[width=14cm]{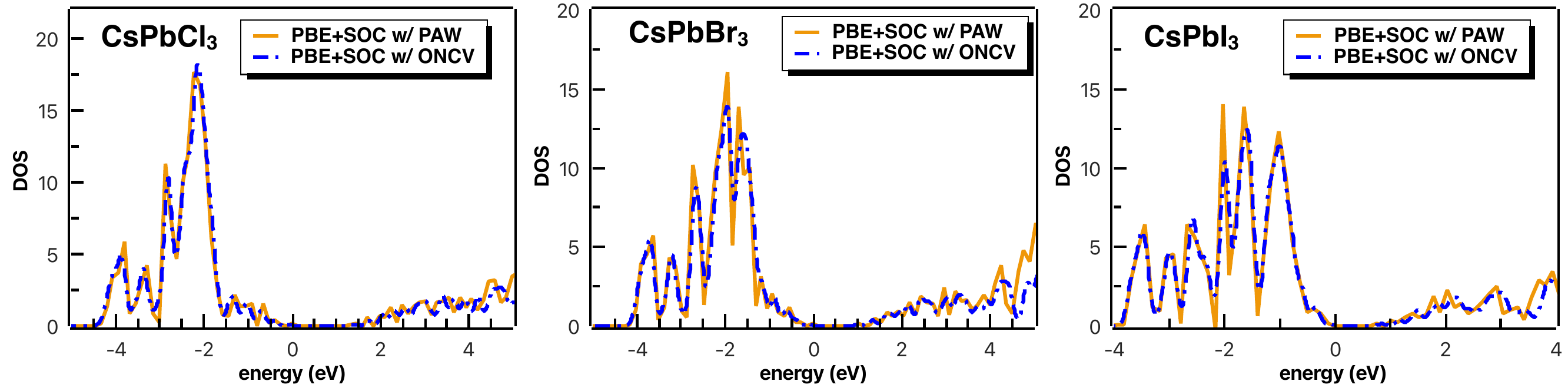}
  \caption{DOS for Pb-containing cubic phase inorganic halide perovskites, i.e., CsPbX$_3$ (X=Cl, Br, I) given by PBE with PAW and PBE with SG15 ONCV pseudopotential, with the SOC effect taken into account.}
  \label{fig:soc_dos}
\end{figure}